\documentclass{aa}  

\usepackage{graphicx}
\usepackage{amsmath}
\usepackage{txfonts}
\usepackage{siunitx}
\usepackage{mathrsfs}
\usepackage{xcolor}
\usepackage{ulem}
\usepackage{dsfont}
\usepackage[labelsep=period]{caption}
\usepackage{subcaption}
\usepackage{nccmath}
\usepackage{mathtools}
\usepackage[inline]{enumitem}
\graphicspath{{./}{figures/}}
\usepackage{hyperref}
\hypersetup{
  unicode=true,      % nonLatin characters in Acrobat's bookmarks
  pdftoolbar=true,    % show Acrobat's toolbar?
  pdfmenubar=true,    % show Acrobat's menu?
  pdffitwindow=true,   % page fit to window when opened
  pdftitle={},
  pdfauthor={Pešta \& Pejcha},
  pdfkeywords={},     % list of keywords
  pdfnewwindow=true,   % links in new window
  colorlinks=true,    % false: boxed links; true: colored links
  linkcolor=blue,     % color of internal links
  citecolor=blue,     % color of links to bibliography
  filecolor=gray,     % color of file links
  urlcolor=blue      % color of external links
}

\newcommand{\teff}{\ensuremath{T_\text{eff}}}
\newcommand{\msun}{\ensuremath{M_\odot}}
\newcommand{\rsun}{\ensuremath{R_\odot}}

\begin{document}

\title{
    Distinguishing between light curves of ellipsoidal variables with massive dark companions, contact binaries, and semidetached binaries using principal component analysis
    }
\titlerunning{
    Lifting the degeneracy of ellipsoidal light curves using PCA
    }

\author{Milan Pe\v{s}ta\thanks{E-mail: \href{mailto:milan.pesta@utf.mff.cuni.cz}{milan.pesta@utf.mff.cuni.cz}} \and Ond\v{r}ej Pejcha}

\authorrunning{Pe\v{s}ta \& Pejcha}

\institute{Institute of Theoretical Physics, Faculty of Mathematics and Physics, Charles University, V Hole\v{s}ovi\v{c}k\'{a}ch 2, Praha 8, 180 00, Czech Republic}

\date{Received 17 August 2024; accepted 21 February 2025}
 
\abstract{
Photometric methods for identifying dark companion binaries -- binary systems hosting quiescent black holes (BHs) and neutron stars (NSs) -- operate by detecting ellipsoidal variations caused by tidal interactions. The limitation of this approach is that contact and semidetached binaries can produce similarly looking light curves. In this work, we address the degeneracy of ellipsoidal light curves by studying the differences between synthetically generated light curves of dark companion, semidetached, and contact binary systems. We inject the light curves with various levels of uncorrelated and correlated Gaussian noise to simulate the effects of instrumental noise and stellar spots. Using principal component analysis (PCA) and Fourier decomposition, we construct low-dimensional representations of the light curves. We find that the first three to five PCA components are sufficient to explain $99\%$ of variance in the data. The PCA representations are generally more informative than the Fourier representation for the same number of coefficients as measured by both the silhouette scores of the representations and the macro recalls of random forest classifiers trained on the representations. The random forest classifiers reach macro recalls from $0.97$ in the complete absence of noise to $0.67$ in the presence of spots and strong instrumental noise, indicating that the classes remain largely separable even under adverse conditions. We find that instrumental noise significantly impacts the class separation only when its standard deviation exceeds $10^{-3}$\,mag, whereas the presence of spots can markedly reduce the class separation even when they contribute as little as $1\%$ of the light curve amplitude. We discuss the application of our method to real ellipsoidal samples, and we show that we can increase the purity of a sample of dark companion candidates by a factor of up to $25$ if we assume a prior purity of $1\%$, significantly improving the cost efficiency of \mbox{follow-up} observations.
}

\maketitle

\section{Introduction} \label{sec:introduction}
Most stellar-mass black holes (BHs) are discovered in binary systems, where their presence is revealed either through high-energy emission from accretion processes \citep[e.g.,][]{Remillard_2006,Corral-Santana_2016} or by gravitational waves radiated during mergers with companions \citep[e.g.,][]{Abbott_2016,Abbott_2023}. Many binaries with neutron stars (NSs) have been discovered in the same way. In reality, only a small fraction of BH and NS binary configurations are expected to yield observable X-ray or gravitational wave signatures. This suggests the existence of a large population of dark companion binaries, which host electromagnetically silent BHs and NSs orbited by normal luminous stars. Given that a significant fraction of BH binaries might actually be wide-orbit binaries \citep{Breivik_2017,Chawla_2022}, characterizing the dark companion population is crucial for enhancing our understanding of the evolution of massive stars and the formation of compact objects.

Without accretion or mergers, the presence of a dark companion in the system can only be infered from subtle photometric, spectroscopic, and astrometric effects that it induces in the companion star. For this reason, only a few dark companion binaries have been discovered so far. A non-exhaustive list of dark companion detections includes two BHs and one BH candidate in the globular cluster NGC 3201 identified using spectroscopy from MUSE \citep{Giesers_2018,Giesers_2019}, one BH or NS identified in data from APOGEE \citep{Thompson_2019}, two BHs found in catalogs of single-lined spectroscopic binaries \citep{Mahy_2022,Shenar_2022}, and three BHs, one NS, and $20$ NS candidates detected in \textit{Gaia} astrometry \citep{Elbadry_2023_sun_like,Elbadry_2023_red_giant,Elbadry_2024_single_NS,Gaia_collaboration_2024,Elbadry_2024_NS_population}. In all these studies, a small number of candidates were selected based on criteria derived from available spectroscopic or astrometric data. The most promising candidates were then followed up with high-resolution spectroscopy, if not already available, to confirm the presence of the dark companion. The limitation of this approach is that spectroscopy and astrometry are available only for a small fraction of stars, significantly reducing the pool of candidates for \mbox{follow-up} analysis. While photometry is available for a much larger number of stars, the challenge lies in identifying the most promising candidates based solely on photometric signatures of dark companions.

In a close binary system consisting of a star and a massive dark companion, the gravitational pull of the companion tidally distorts the star, inducing ellipsoidal variations in the light curve of the system. In principle, by sifting through large photometric surveys and identifying stars that exhibit ellipsoidal variations, we can select candidates for \mbox{follow-up} analysis that are likely to harbor dark companions. Recent examples of such work include \citet{Green_2023}, who identified over $15\,000$ ellipsoidal variables in data from TESS, \citet{Gomel_2023}, who presented over $6\,000$ dark companion candidates from \textit{Gaia} DR3, and \citet{Gomel_2021_ogle}, who studied over $10\,000$ ellipsoidal variables from {OGLE}. The problem with this method is that besides dark companion binaries, ellipsoidal samples typically contain large numbers of contact binaries, semidetached binaries, and possibly other types of objects that produce similar light curves. In fact, dark companion binaries most likely make up only a small fraction of ellipsoidal variables, making it extremely cost-inefficient to follow up on all candidates with high-resolution spectroscopy.

To increase the fraction of dark companion binaries in ellipsoidal samples, further filtering is required. For example, we could filter objects based on the quality of their spectral energy distribution fits assuming a single-star model \citep{Kapusta_2024} or we could consider only objects with high binary mass functions \citep{Rowan_2024}. However, the former method requires multiband photometry, which is not always available, while the latter relies on radial velocity measurements, thus defeating the goal of avoiding the need for spectroscopy in the candidate selection process. In our previous work \citep{Pesta_2023}, we used Bayesian mixture modeling to isolate a sample of contact binaries from the Kepler Eclipsing Binary Catalog. In principle, we could use the same method to exclude contact binaries from samples of ellipsoidal variables, but the method requires estimates of effective temperatures and luminosities for all objects in the sample, limiting its applicability.

A particularly attractive way of selecting dark companion binary candidates using only information contained in their broadband photometric light curves was developed by \citet{Gomel_2021_mmm_ratio,Gomel_2021_amplitude}, who introduced a proxy for the minimum mass ratio of dark companion binaries derived from the observed ellipsoidal amplitude of the system. This proxy, which they termed the modified minimum mass ratio (mMMR), is always strictly lower than the actual mass ratio of the system, and its large values can be indicative of the presence of a massive dark companion. The method is most sensitive to dark companion binaries with primaries close to filling their Roche lobes and inclinations close to $90^\circ$. Conversely, low-inclination systems or systems with a primary that did not yet evolve to fill its Roche lobe will show small mMMR even for large mass ratios. The method assumes that and all variability comes from the tidal deformation of the primary induced by the dark companion. When this assumption is violated, the method yields spurious results, resulting in high false-positive rates. For example, \citet{Nagarajan_2023} spectroscopically followed up on the $14$ most promising candidates obtained using the mMMR method by \citet{Gomel_2023} and found that all harbor a low-mass non-degenerate star instead of a dark companion, with spotted contact binaries being the most likely culprits behind the false positives. Consequently, the efficiency of the mMMR method hinges on the purity of the ellipsoidal sample, which is typically low due to the prominent presence of contaminants \citep[e.g.,][]{Green_2023}.

A proper way to address the issue of photometric identification of dark companion binaries would be to train a machine learning classifier on a large sample of ellipsoidal light curves for which we know the true nature of the systems, allowing the classifier to learn the differences between the classes and automatically detect dark companion binaries in new data. Many have followed this approach in the wider context of automatic classification of periodic variables \citep[e.g.,][etc.]{Paczynski_2006,Pawlak_2016,Soszynski_2016,Jayasinghe_2019,Cheung_2021}. However, this method requires a \mbox{well-curated} training sample in which all classes are sufficiently represented. This is generally not an issue in variable star classification, where large samples of studied objects are readily available, but a representative sample of confirmed dark companion binaries is currently lacking, preventing us from following this approach in the context of dark companion binary identification.

Even in the absence of a well-defined sample of dark companion binaries, it is still possible to study the degeneracy of ellipsoidal light curves using theoretical models. In this work, we investigate the similarities and differences between synthetically generated light curves of dark companion binaries, semidetached binaries, and contact binaries, which we consider to be to the most likely dark companion binary impostors. We inject the light curves with various levels of uncorrelated and correlated Gaussian noise to simulate the effects of instrumental noise and stellar spots (Sect.~\ref{sec:data}), allowing us to study the separation of the classes under adverse observing conditions. To better visualize the light curves and make the differences between the classes more pronounced, we reduce the light curves using principal component analysis (PCA) and Fourier decomposition. We compare the informativeness of the PCA and Fourier representations using the silhouette score, and we quantify the separation of the classes in each representation using the macro recall of random forest classifiers trained on the representations. We describe the methodological details of our analysis in Sect.~\ref{sec:methods}, and we present the results of our study in Sect.~\ref{sec:results}. We summarize and discuss the implications of our findings in Sect.~\ref{sec:conclusions}.

\section{Synthetic data} \label{sec:data}
The small number of confirmed dark companion binaries prevents us from using real observations to systematically study the degeneracy of ellipsoidal light curves. To overcome this limitation, we generated synthetic light curves of dark companion binaries and their common contaminants (Sect.~\ref{sec:physical_models}), which we further modified with correlated and uncorrelated noise to account for the effects of instrumental noise and stellar spots (Sec.~\ref{sec:noise_oversampling}).

\subsection{Physical models}
\label{sec:physical_models}
We used PHOEBE v2.4.10 \citep{Prsa_2016,Conroy_2020} to generate synthetic light curves of dark companion binaries, semidetached binaries, and contact binaries. We started by initializing the default detached, semidetached or contact binary system, conditional on the type of the variable we wanted to generate. In all cases, we set the passband to \texttt{TESS:T}, the number of triangles of the stellar components to $10\,000$, and we kept the default limb darkening calculation settings, with the coefficients interpolated directly from either the \texttt{PHOENIX} or the \texttt{ck2004} model atmosphere tables, depending on the effective temperatures of the stellar components. In dark companion systems, we set \texttt{distortion\_method}~$=$~\texttt{none} for the dark companion, allowing us to isolate the variations caused by the tidal deformation of the star without accounting for the presence of eclipses. In all other cases, we kept the default \texttt{distortion\_method}~$=$~\texttt{roche}. 

For each binary class, we defined a grid of physical and orbital parameters that affect the shape of the light curve. In the case of dark companion binaries, the light curve does not significantly depend on the mass $M$ nor the effective temperature $\teff$ of the stellar component but rather on the mass ratio $q$, the inclination $i$, and the semimajor axis $a$ of the system. We therefore fixed $M$~$=$~$1\,\msun$ and $\teff$~$=$~$6\,000$\,K. We varied $q$ from $0.05$ to $10$ with a step of $0.05$ for $0.05 \le q \le 1$ and a step of $1$ for $1$~$<$~$q$~$\le$~$10$, $i$ from $5^\circ$ to $90^\circ$ with a step of $5^\circ$, and $a$ from $1\,\rsun$ to $10\,\rsun$ with ten evenly spaced steps in the logarithmic scale. We fixed the equivalent radius $R$ of the stellar component at $1\,\rsun$. Due to the scaling properties of the Roche potential, varying $R$ has the same impact on the shape of the light curve as varying $a$.

Compared to the dark companion case, the light curves of semidetached variables additionally depend on the ratios of $\teff$ and $R$ of the stellar components. We considered $\teff$ ratios of $0.5$, $1$, and $2$, and we varied $\teff$, the gravity brightening coefficients, and the bolometric reflection coefficients of the components accordingly so that both the primary and the secondary were covered by the available atmosphere tables and the system passed all PHOEBE internal checks. While the chosen $\teff$ ratio grid is rather coarse, it is sufficient for our purposes as we expect the greatest degeneracy with the light curves of dark companion binaries and contact binaries when the components in semidetached systems have similar effective temperatures. Additionally, we considered four different $R$ ratios: $0.1$, $0.5$, $2$, and $5$, with $R$ of the primary fixed by the condition that the primary fills its Roche lobe. We sampled $q$, $i$, and $a$ in the same way as in the dark companion case.

The simplest is the case of contact binary stars, whose light curves depend primarily on $q$, $i$, and the fill-out factor $f$ of the system. We sampled $i$ and $q$ in the same way as in the previous cases, with the only difference being that we limited $q$ to $0.05$~$\le$~$q$~$\le$~$1$. We considered $f$~$=$~$0.15$, $0.25$, $0.5$, and $0.75$, and we assumed that the two components share a common atmosphere with $\teff$~$=$~$6000$\,K. We present a summary of the parameter ranges used to generate the synthetic light curves in Table~\ref{tab:parameter_ranges}.

Not all parameter combinations produced a valid light curve. Some setups were not covered by either atmosphere table or yielded a configuration that was incompatible with the assumed binary class (e.g., Roche overflow in dark companion binaries). We excluded these configurations from the analysis. We also excluded any setup that yielded a light curve with a photometric amplitude smaller than $0.01$ mag. By performing these cuts, we obtained samples of $1\,122$ dark companion, $37\,386$ semidetached, and $1\,302$ contact binary synthetic light curves. The observed disparity in the sample sizes does not reflect the relative occurrence rates of the three binary classes but rather the relative extents of their parameter spaces. To counter this imbalance, we randomly undersampled the semidetached binary class by a factor of $20$, resulting in a total of $1\,846$ semidetached light curves. Finally, we randomly split the data into training, validation, and test sets, with $20\%$ of each variable class going to the test set and $20\%$ going to the validation set.

\begin{table*}
    \caption{Ranges of physical and orbital parameters used to generate synthetic light curves of dark companion, semidetached, and contact binary systems.}
    \label{tab:parameter_ranges}
    \begin{center}
    \begin{tabular}{cccc}
    \hline\hline
    Parameter & Dark companion binaries  & Semidetached binaries & Contact binaries\\
    \hline
    $M_{1}$ ($\msun$) & $1$ & $1$ & $1$\\
    $q$ & $0.05$--$1$ ($\Delta q = 0.05$) + $1$--$10$ ($\Delta q = 1$) & $0.05$--$1$ ($\Delta q = 0.05$) + $1$--$10$ ($\Delta q = 1$) & $0.05$--$1$ ($\Delta q = 0.05)$\\
    $T_{\text{eff},1}$ (K) & $6\,000$ & $6\,000$, $8\,000$, $15\,000$ & $6\,000$\\
    $T_{\text{eff},2}/T_{\text{eff},1}$ & -- & $0.5$, $1$, $2$ & 1\\
    $R_{1}$ ($\rsun$) & $1$ & -- & --\\
    $R_{2}/R_{1}$ & -- & $0.1$, $0.5$, $2$, $5$ & --\\
    $i$ ($^\circ$) & $5$--$90$ ($\Delta i = 5^\circ$) & $5$--$90$ ($\Delta i = 5^\circ$) & $5$--$90$ ($\Delta i = 5^\circ$)\\
    $\log{a/\rsun}$ & $0$--$1$ ($\Delta \log{a/\rsun}=0.11$) & $0$--$1$ ($\Delta \log{a/\rsun}=0.11$) & --\\
    $f$ & -- & -- & $0.15$, $0.25$, $0.5$, $0.75$\\
    \hline
    \end{tabular}
    \tablefoot{Parameters denoted with subscripts $1$ and $2$ refer to the primary and secondary components, respectively.}
    \end{center}
    \end{table*}

We generated the light curves in the magnitude space with a resolution of $100$ points per orbit, covering phase from $-0.5$ to $0.5$. We shifted the light curves so that the phase $0$ corresponds to the primary minimum, and we discarded the rightmost point of each light curve to avoid redundancy at phase $0.5$. We normalized the light curves by vertically shifting and rescaling them in such a way that the maximum and the minimum were equal to $1$ and $0$, respectively. Hereafter, we shall refer to this sample of normalized synthetic light curves as S0.

\subsection{Addition of noise and oversampling} \label{sec:noise_oversampling}
To account for the effects of instrumental noise and stellar spots, we injected the sample S0 with various levels of uncorrelated and correlated noise. We modeled the instrumental noise as uncorrelated Gaussian noise with standard deviations $\sigma_\text{WN}$~$=$~$10^{-4}$, $10^{-3}$, and $10^{-2}$\,mag, covering almost the entire range of the TESS noise characteristic curve \citep{Ricker_2015}. We modeled the effects of spots as correlated Gaussian noise, which we generated using the \texttt{scikit-learn} implementation of Gaussian processes with a periodic \texttt{ExpSineSquared} kernel. We considered correlated noise with standard deviations $\sigma_\text{CN}$~$=$~$0.01$, $0.05$, and $0.10$ of the unperturbed light curve amplitude and correlation length scales $l_\text{CN}$~$=$~$0.25$, $0.50$, and $1.00$ of the orbital period. The standard deviation of the injected correlated noise is proportional to the amplitude of the unperturbed light curves, because we want to simulate a scenario in which stellar spots account for a specific fraction of the overall light curve amplitude, having the same relative effect on all light curves. Also, since we assume that all variability in the light curves before noise injection comes from eclipses and ellipsoidal variations, the light curve amplitude should be zero for a system observed exactly face-on irrespective of the presence of spots, which would not be the case if we injected correlated noise with an absolute standard deviation.

For each combination of the levels of correlated and uncorrelated noise, including the complete absence of noise, we generated $10$ realizations of each light curve in the sample S0, resulting in $40$ synthetic datasets with well-sampled noise distributions (Table~\ref{tab:synthetic_samples}). To prevent data leakage, we oversampled the training, validation, and test sets separately, so that each light curve and its noisy realizations were present in only one of the sets. Each synthetic sample contains a total of $42\,700$ light curves, with $11\,220$ light curves coming from the dark companion class, $18\,460$ from the semidetached class, and $13\,020$ from the contact class.

\begin{figure*}
    \centering
    \includegraphics[width=0.9\textwidth]{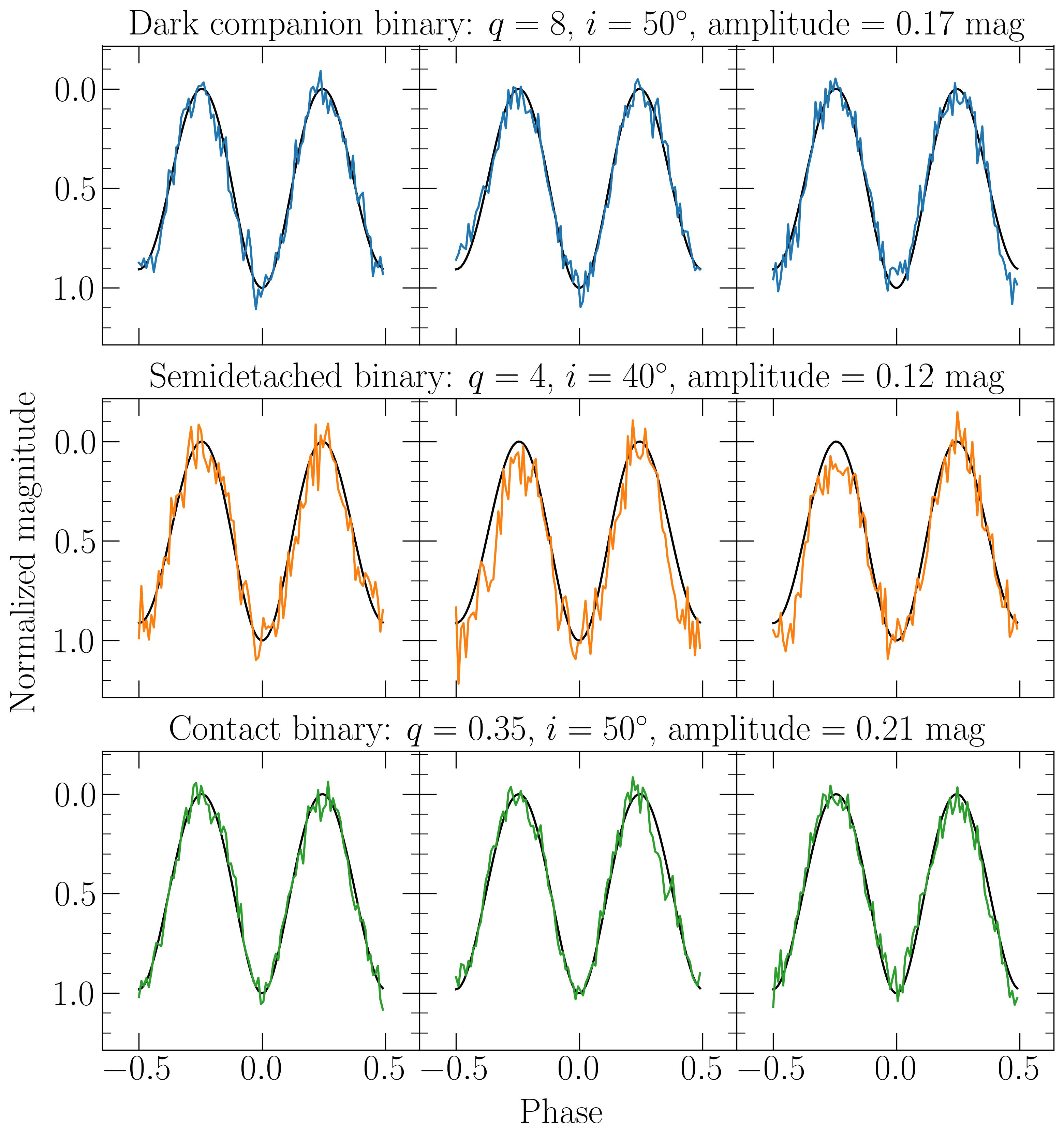}
    \caption{Examples of normalized synthetic light curves of dark companion (top panel), semidetached (middle panel), and contact binary systems (bottom panel). Each panel contains three subplots showing different light curve realizations of the system characterized by the mass ratio $q$ and inclination $i$ specified at the top of the panel. The solid black lines represent the noise-free light curves of the systems, while the colored lines show the light curves perturbed with noise. All noisy light curves were injected with uncorrelated Gaussian noise with a standard deviation $\sigma_\text{WN} = 10^{-2}$\,mag and correlated Gaussian noise modeled using the \texttt{ExpSineSquared} kernel with a standard deviation $\sigma_\text{CN} = 0.05$ of the unperturbed light curve amplitude (given at the top of each panel) and correlation length scale $l_\text{CN} = 0.25$ of the orbital period.}
    \label{fig:light_curve_examples}
\end{figure*}

Finally, after injecting the light curves with noise, we normalized them by: i) fitting each light curve with a fourth-order Fourier series, ii) horizontally shifting the light curves so that the primary minimum (corresponding to the maximum magnitude) of the Fourier fit is at phase $0$, iii) vertically shifting and rescaling the light curves so that the Fourier fit has a minimum and maximum of $0$ and $1$, respectively. We show examples of normalized synthetic light curves of dark companion, semidetached, and contact binaries in Fig.~\ref{fig:light_curve_examples}.

\section{Methods} \label{sec:methods}
Light curves can be viewed as vectors in a high-dimensional space, with the dimension of the space given by the total number of data points in the light curve. Current space-based photometric surveys have typical sampling frequencies on the order of seconds to minutes, resulting in densely sampled light curves with thousands of points. Intuitively, the denser the sampling, the more information the light curve contains and the easier it should be to construct a classifier that can distinguish between different types of objects generating the light curves. In practice, the high-dimensional nature of the data might actually hurt the performance of the classifier. The reason is that as the dimension of the data increases, the number of samples required to evenly cover the space grows exponentially, and the available data become increasingly sparse. This phenomenon, known as the curse of dimensionality \citep{Bellman_1957}, is further exacerbated by the fact that real-life light curves are often contaminated by noise and outliers, which in combination with overfitting can lead to poor generalization to previously unseen data.

Training an accurate light curve classifier that generalizes well to new observations requires a training sample that is representative of the true distribution of the data. However, in our synthetic dataset, the relative frequencies of the binary classes and their within-class parameter distributions are the result of our choices and do not reflect the actual occurrence rates of the different binary configurations. In addition, we injected the synthetic data with the correlated and uncorrelated noise to systematically study the separation of the classes under various noise levels rather than to simulate the noise characteristics of real data, which are specific to the instrument and observing conditions.

Despite the synthetic data not being representative, we can still use it to quantify our ability to distinguish between the three binary classes. This can be achieved by constructing a discriminative low-dimensional representation of the data and investigating the separation of the classes in this representation. The idea is based on the observation that classification tasks often include a dimensionality reduction preprocessing step where the data is projected to a low-dimensional space in a way that preserves most of the information contained in the original high-dimensional data. In the absence of a representative training sample, we can separate dimensionality reduction from the classification task, allowing us to focus on the quality of the data representation rather than on the performance of the classifier. Apart from alleviating the effects of the curse of dimensionality and overfitting, dimensionality reduction also makes data easier to visualize and interpret, resulting in a more compact and informative representation. Once we have collected a representative sample, we can project the data to the learned low-dimensional space and train a classifier on the reduced data, ensuring robust generalization to new observations.

There are many dimensionality reduction methods, ranging in complexity from simple summary statistics and direct encodings to sophisticated latent representations learned directly from the data through optimization. In this work, we utilized PCA \citep{Pearson_1901,Hotelling_1933}, a simple linear method that is easy to interpret and requires next to no hyperparameter tuning. By performing PCA separately on the three binary classes, we obtained three distinct latent representations of the synthetic data, each optimized to capture the underlying structure of the respective class (Sect.~\ref{sec:pca_representations}). As a baseline for comparison, we also expanded the light curves into Fourier coefficients, which is a standard practice in time series analysis (Sect.~\ref{sec:fourier_representation}). We used two metrics to quantify the separation of the classes in the PCA and Fourier representations: the silhouette score, which compares the average intra-class distance to the average inter-class distance (Sect.~\ref{sec:silhouette_score_methods}), and the macro recall of random forest classifiers, which can be interpreted as a measure of the mean non-overlap of the classes in the feature space (Sect.~\ref{sec:random_forests_macro_recall_methods}).

\subsection{PCA representations} \label{sec:pca_representations}
Our motivation for using PCA in this work is multifold: 
\begin{enumerate*}[label=(\roman*)]
    \item PCA is a well-tested, widely adopted method that is easy to use and computationally very efficient.
    \item PCA is a simple yet powerful enough method to gain intuition with dimensionality reduction, allowing us to illustrate the idea of using data-driven methods for feature extraction and classification before moving on to more sophisticated methods.
    \item With only one tunable hyperparameter, the number of retained principal components, PCA does not require extensive tuning, making it an ideal starting point in any dimensionality reduction task.
    \item The linear character of PCA representations makes them robust to noise, meaning that small perturbations in light curves do not significantly alter their representations. Consequently, we can perform PCA on noiseless light curves and then project noisy light curves using the obtained principal components, knowing that similar light curves will have similar representations, which is not necessarily the case with nonlinear methods.
    \item When interpreted in the original magnitude space, the principal components can be interpolated, yielding a set of continuous functions that are orthogonal under the standard $L^2$ inner product. In analogy with Fourier decomposition, these continuous principal components can then be used to generalize the PCA representation to light curves with arbitrary sampling.
\end{enumerate*}

In our analysis of the synthetic light curves, we utilized the \texttt{scikit-learn} implementation of PCA. The implementation returns the unit eigenvectors and the eigenvalues of the covariance matrix of the centered data, meaning that the mean is subtracted from each vector before the computation. Using the noiseless sample S0 as input, we performed PCA separately on the light curves of each binary class, yielding three distinct orthonormal bases of principal components. The synthetic light curves have a resolution of $100$ points per orbit, with the last point removed for reasons related to the point (v) above, resulting in vectors of length $N_\text{grid}$~$=$~$99$. PCA preserves the dimensionality of the feature space, so the number of principal components in each basis is also $N_\text{grid}$.

We performed PCA on the normalized light curves instead of the original light curves in the magnitude space to ensure that the principal components reflect the intrinsic variations in the shapes of the light curves rather than the variations in amplitude. Since the amplitudes vary significantly within the classes, performing PCA on the original light curves would yield principal components that are dominated by amplitude, thereby obscuring the shape variations and resulting in suboptimal data representation. By factoring out the absolute scale before performing PCA and treating amplitude as a separate feature, we ensure that the morphology of the light curves is properly captured by the principal components, maximizing the overall information content of the representation.

If we denote the $i$th principal component of the class $K$ as $\mathbf{\text{e}}^{K}_i$, where $K$~$=$~DC, SD, C for the dark companion, semidetached, and contact binary classes, respectively, we can expand any normalized light curve $\mathbf{v}$ evaluated on the same grid as the principal components as
\begin{ceqn}
\begin{equation}
    \mathbf{v} = \mathbf{\text{e}}^{K}_0 + \sum_{j=1}^{N_\text{grid}} c^{K}_{j} \mathbf{\text{e}}^{K}_j,
    \label{eq:pca}
\end{equation}
\end{ceqn}
where $\mathbf{\text{e}}^{K}_0$ is the mean normalized light curve of the class $K$ and the vector of coefficients $\mathbf{c}^{K}_{N_\text{grid}}$~$=$~$(c^{K}_{1}, c^{K}_{2}, \ldots, c^K_{N_\text{grid}})$ gives the coordinates of $\mathbf{v}$ in the PCA basis of the class $K$. By mean normalized light curve of the class $K$ we mean a light curve obtained by taking the average of all normalized light curves in the class at each phase point of the grid. If we consider the full vector of coefficients, the representation is perfect with no information loss. If we keep only the first $n$~$<$~$N_\text{grid}$ coefficients, we can write the reconstruction of $\mathbf{v}$ as
\begin{ceqn}
\begin{equation}
    \mathbf{v}^{K}_{n} = \mathbf{\text{e}}^{K}_0 + \sum_{j=1}^{n} c^{K}_{j} \mathbf{\text{e}}^{K}_j,
    \label{eq:pca_reduced}
\end{equation}
\end{ceqn}
with the reduced vector of coefficients $\mathbf{c}^{K}_{n}$~$=$~$(c^{K}_{1}, c^{K}_{2}, \ldots, c^{K}_{n})$ constituting the $n$-dimensional PCA representation of the light curve in the basis of the class $K$. The superscript $K$ on the left side of Eq.~(\ref{eq:pca_reduced}) emphasizes that the light curve reconstructed from the first $n$ principal components is class-dependent, unlike the fully reconstructed light curve in Eq.~(\ref{eq:pca}).

The coefficients $\mathbf{c}^{K}_{n}$ can be obtained either by projecting the light curve onto the principal components or equivalently by fitting the light curve with a linear combination of the principal components using least squares. The equivalence of the two methods allows us to easily generalize the PCA representation to light curves with arbitrary sampling by interpolating the principal components in the normalized magnitude space and fitting the light curve with the interpolated principal components. By allowing arbitrary nonuniform sampling, we are no longer guaranteed the orthogonality of the principal components when evaluated on the new grid. Consequently, the PCA coefficients obtained from least squares fitting can change as we increase the number of the components in the fit. However, for densely sampled light curves, we expect the most informative coefficients to converge for $n$~$\ll$~$N_\text{grid}$, yielding a representation that is robust to the light curve sampling. With a typical sampling frequency of the current space-based photometric surveys on the order of seconds to minutes, this condition is satisfied for a vast majority of light curves.

There is one issue with defining the PCA representation using principal components with a unit norm. If we doubled the resolution of the synthetic light curves and performed PCA on this finer grid, the photometric amplitude of the principal components would be approximately $\sqrt{2}$ times smaller than the amplitude of the original principal components. This is because the principal components are normalized to have a unit norm, and the finer grid contains twice as many points as the original grid. To avoid this issue, we fixed the scaling of the principal components to have a unit amplitude in the normalized magnitude space. This scaling ensures that the PCA coefficients are directly comparable between representations obtained from grids with different resolutions. Denoting the rescaled principal components as $\mathbf{\tilde{\text{e}}}^{K}_j$, we can write the projection of the light curve $\mathbf{v}$ onto the first $n$ rescaled principal components of the class $K$ as
\begin{ceqn}
\begin{equation}
    \mathbf{v}^{K}_{n} = \mathbf{\text{e}}^{K}_0 + \sum_{j=1}^{n}\tilde{c}^{K}_{j} \mathbf{\tilde{\text{e}}}^{K}_j,
    \label{eq:pca_rescaled}
\end{equation}
\end{ceqn}
where the vector of rescaled coefficients $\mathbf{\tilde{c}}^{K}_{n}$~$=$~$(\tilde{c}^{K}_{1}, \tilde{c}^{K}_{2}, \ldots, \tilde{c}^{K}_{n})$ forms the $n$-dimensional PCA representation of $\mathbf{v}$ in the rescaled PCA basis of the class $K$. The generalization of the rescaled PCA representation to light curves with arbitrary sampling is achieved in the same way as in the case of the original unit PCA representation.

Both the unit and the rescaled PCA representations operate on normalized light curves, which are vertically shifted and rescaled so that the amplitude of the fourth-order Fourier fit is equal to unity. By normalizing the light curves, we lose the information about their absolute scaling. To recover this information, we prefix the vector of PCA coefficients in both representations with the amplitude obtained from the Fourier fit, increasing the dimensionality of the representations by one. We shall denote the amplitude as $c_0$ and the extended unit and rescaled PCA representations as capital $\mathbf{C}^{K}_{n}$~$=$~$(c_0, c^{K}_{1}, c^{K}_{2}, \ldots, c^{K}_{n})$ and $\mathbf{\tilde{C}}^{K}_{n}$~$=$~$(c_0, \tilde{c}^{K}_{1}, \tilde{c}^{K}_{2}, \ldots, \tilde{c}^{K}_{n})$, respectively. This way, the amplitude of the light curve is encoded as the zeroth element of the extended PCA representations, allowing us to rescale the normalized light curve back to the original magnitude space by multiplying the PCA coefficients and the mean light curve with $c_0$.

\subsection{Fourier representation}
\label{sec:fourier_representation}
Historically, expansion to Fourier coefficients has been the most popular method for dimensionality reduction of time series data. In discrete Fourier series, we decompose a uniformly sampled normalized light curve $\mathbf{v}$ of length $N_\text{grid}$ into a linear combination of harmonics of increasing order up to the Nyquist frequency, totalling $N_\text{grid}$ coefficients. For an odd $N_\text{grid}$, this can be expressed as
\begin{ceqn}
\begin{equation}
    \mathbf{v} = a_0{\mathds{1}} + \sum_{j=1}^{(N_\text{grid}-1)/2} a_j \mathbf{cos}_j + b_j \mathbf{sin}_j,
    \label{eq:fourier}
\end{equation}
\end{ceqn}
where $\mathds{1}$ is a constant $N_\text{grid}$-dimensional vector of ones and $\mathbf{cos}_j$ and $\mathbf{sin}_j$ are vectors of the $j$th-order cosine and sine harmonics sampled on the same grid as the light curve. The base period of the harmonics is given by the period of the light curve, which is equal to one for phased light curves.

To emphasize the similarity between Fourier decomposition and PCA, we can factor out the mean light curve and express $\mathbf{v}$ as
\begin{ceqn}
    \begin{equation}
        \mathbf{v} = \mathbf{e}^K_0 + \sum_{j=1}^{N_\text{grid}} \tilde{c}^\text{F}_j \mathbf{\tilde{e}}^\text{F}_j,
        \label{eq:fourier_unified}
    \end{equation}
\end{ceqn}
where $\mathbf{e}^K_0$ is the mean normalized light curve of the class $K$ and
\begin{ceqn}
    \begin{equation}
        \mathbf{\tilde{e}}^\text{F}_j = \begin{cases}
        \mathds{1} & \text{if } j = 1, \\
        \mathbf{cos}_{j/2} & \text{if } j>1 \text{ is even}, \\
        \mathbf{sin}_{(j-1)/2} & \text{if } j>1 \text{ is odd}.
        \end{cases}\\
        \label{eq:fourier_basis}
    \end{equation}
\end{ceqn}
We use the tilde notation from the previous section to emphasize the fixed scaling of the discretized Fourier basis elements. The mean light curve $\mathbf{e}^K_0$ can belong to any class $K$~$=$~DC, SD, C. In this work, we are mainly interested in searching for dark companion binaries, so we choose $K$~$=$~DC. By keeping only the first $n$ coefficients, we can reduce the dimensionality of the data while capturing the information about frequencies up to the harmonic preceded by the $n$th coefficient. We write the reconstruction of $\mathbf{v}$ using the first $n$ coefficients as
\begin{ceqn}
    \begin{equation}
        \mathbf{v}^{F}_{n} = \mathbf{e}^\text{DC}_0 + \sum_{j=1}^{n} \tilde{c}^\text{F}_j \mathbf{\tilde{e}}^\text{F}_j.
        \label{eq:fourier_unified_n_coeffs}
    \end{equation}
    \end{ceqn}
Motivated by the analogy between Eqs.~(\ref{eq:fourier_unified_n_coeffs}) and (\ref{eq:pca_reduced}), we define the $n$-dimensional Fourier representation of $\mathbf{v}$ as the vector of coefficients $\mathbf{\tilde{c}}^F$~$=$~$(\tilde{c}^\text{F}_{1}, \tilde{c}^\text{F}_{2}, \ldots, \tilde{c}^\text{F}_{n})$. Consequently, all the considerations from the previous section regarding the generalization of the PCA representations to light curves with arbitrary samplings apply to the Fourier representation as well.

In addition to the standard ``rescaled'' Fourier representation (the discretized harmonics have fixed scaling), we also define the unit Fourier representation $\mathbf{c}^\text{F}_{n}$~$=$~$(c^\text{F}_{1}, c^\text{F}_{2}, \ldots, c^\text{F}_{n})$, where the coefficients are obtained with respect to the normalized Fourier basis elements with unit norms. By analogy with the PCA representations, we define the extended Fourier representation as $\mathbf{\tilde{C}}^{F}_{n}$~$=$~$(c_0, \tilde{c}^{F}_{1}, \tilde{c}^{F}_{2}, \ldots, \tilde{c}^{F}_{n})$, where $c_0$ is the amplitude of the light curve defined in the previous section. The extended unit Fourier representation is defined analogously as $\mathbf{C}^{F}_{n}$~$=$~$(c_0, c^{F}_{1}, c^{F}_{2}, \ldots, c^{F}_{n})$.

Hereafter, we collectively refer to the PCA and Fourier representations as the latent representations, and we refer to the vector spaces spanned by the coefficients of the latent representations as the latent spaces. We further distinguish between rescaled and unit latent representations, which differ in the scaling of the basis vectors. The extended latent representations, be they rescaled or unit, include the amplitude of the light curves as the zeroth element, ensuring that the information about the absolute scale is preserved. We omit the lower index $n$ in the notation when we refer to the latent representations in general, without reference to a specific dimension.

\subsection{Silhouette score}
\label{sec:silhouette_score_methods}
To compare the class separation in the PCA and Fourier representations, we calculated the weighted silhouette scores of the representations $\mathbf{c}^\text{DC}_{n}$, $\mathbf{c}^\text{SD}_{n}$, $\mathbf{c}^\text{C}_{n}$, and $\mathbf{c}^\text{F}_{n}$ as a function of the dimension of the representation $n$~$=$~$1\text{--}9$. We provide the definition of the silhouette score and the details of the calculation in Appendix \ref{app:silhouette_score}.

We evaluated the silhouette scores on the unit representations instead of the extended unit representations, because we wanted to maximize the class separation with respect to the shapes of the light curves independent of their amplitudes. The amplitude is a robust discriminative feature and, in the presence of strong noise, it can skew the silhouette scores of low-dimensional representations toward artificially high values, potentially obscuring the true number of coefficients that maximize the separation of the classes. Also, the amplitude affects the silhouette scores of all latent representations in roughly the same way, and since we are only interested in the difference between the silhouette scores of different representations and not their absolute values, we can safely omit the amplitude from the calculation. To provide a baseline, we also calculated the silhouette score of the full representation consisting of \mbox{$99$-dimensional} vectors of normalized light curves in the magnitude space. We performed the calculation on the validation sets of the samples W0C0, W100C0, W0C10L50, and W100C10L50, which are the synthetic samples with the lowest and the highest levels of uncorrelated and correlated noise, either separately or in combination. Hereafter, we shall refer to these samples as the corner cases. The corner cases provide the most extreme conditions for the separation of the classes, and the conclusions drawn from them are generally applicable to the intermediate cases as well.

\subsection{Macro recall and random forest classifiers}
\label{sec:random_forests_macro_recall_methods}
There are several downsides to using the silhouette score as a measure of class separation in the latent space. First, due to the silhouette score not being invariant under independent rescaling of the features, the absolute value of the silhouette score is meaningless, only the difference between the silhouette scores of different representations is informative. Second, we calculated the silhouette scores as a function of the number of coefficients in the representation. However, not all coefficients are equally informative, meaning that the first $n$ coefficients can yield a lower silhouette score that the same number of non-consecutive but more informative coefficients. Third, the silhouette score as a measure of separation is best suited for convex clusters, which is not necessarily the case for the binary classes in the latent representations, not to mention the full representation. For concave overlapping and/or nested clusters, the silhouette score can be close to zero even if the clusters are perfectly separated. For these reasons, we turn to a more robust measure of class separation: the macro recall, which is a metric that is commonly used to assess the performance of classifiers in multi-class classification tasks. We provide the definition of macro recall and details of its calculation in Appendix \ref{app:macro_recall}.

One limitation of the macro recall is that it is specific to the classifier, which means that its value can change when we evaluate it for a different classifier on the same data or when we use the same classifier with different hyperparameters. Choosing the optimal classifier that yields the best macro recall is a nontrivial task that requires hyperparameter tuning and cross-validation to select the best performing model. Without any prior knowledge of the problem, the best approach is to start with a simple and robust classifier that does not require excessive tuning, such as the random forest classifier \citep{Breiman_2001}, which is known to perform well on a wide range of problems and is not too sensitive to the choice of hyperparameters.

In our analysis, we used the \texttt{scikit-learn} implementation of the random forest classifier. For each synthetic sample (Table~\ref{tab:synthetic_samples}), we trained a number of random forest classifiers with different hyperparameter configurations on the extended rescaled representations $\mathbf{\tilde{C}}^{K}_{n}$, $K$~$=$~DC, SD, C, F and $n$~$=$~$1\text{--}9$. To provide a baseline, we also considered the ($1+99$)-dimensional extended full representation consisting of photometric amplitudes + normalized light curves, and the one-dimensional representation consisting of photometric amplitudes only. In addition, we augmented each representation except the extended full representation with the variances of the coefficients obtained from the least squares fits of the photometric amplitude and the latent coefficients, and we retrained the random forest classifiers on the augmented representations. We did this to examine whether the uncertainties of the coefficients contain useful information that could help us better separate the classes. We calculated the uncertainty of the photometric amplitude as the variance of the residuals from the fourth-order Fourier fit of the light curve. Given an extended rescaled representation $\mathbf{\tilde{C}}^{K}_{n}$, we denote its augmented version as $\mathbf{\tilde{C}}^{K+\text{V}}_{n}$, where $K$~$=$~DC, SD, C, F, and $n$ is the number of coefficients in the representation.

We trained the random forest classifiers on the extended rescaled representations to ensure that the absolute scale of the light curves is taken into account when separating the classes. We did not include the amplitude in the calculation of the silhouette scores, because it could bias the optimal number of coefficients toward lower values in the presence of strong noise, but the properties of the random forest classifier make it possible to include the amplitude in the input without obscuring the discriminative patterns in the coefficients of the representations. In addition, the macro recall is an absolute measure of class non-overlap, which means that we are actually interested in its values, not just its differences between different representations. To achieve the best possible macro recall, it is necessary we consider all the information contained in the light curves, including the amplitude.

We obtained the optimal hyperparameters of the random forest classifiers trained on each representation of each synthetic sample by performing a basic grid search for selected hyperparameters. Namely, we considered: the number of trees in the forest \texttt{n\_estimators}~$=$~$100$, $500$; the minimum number of samples required to be at a leaf node \texttt{min\_samples\_leaf}~$=$~$1$, $10$; and the method for selecting the number of features at each split \texttt{max\_features}~$=$~\texttt{sqrt}, \texttt{log2}, \texttt{None}. We used the default values for the remaining hyperparameters. In all cases, we trained the random forest classifiers on the training sets, performed the hyperparameter tuning on the validation sets, and evaluated the best performing classifiers on the test sets of the synthetic samples.

\section{Results} \label{sec:results}
In this section, we present the results of our analysis of the synthetic light curves of dark companion, semidetached, and contact binary systems. In Sect.~\ref{sec:pca_models}, we provide an overview of the PCA models of the three binary classes. We visually inspect the PCA and Fourier representations of the light curves in Sect.~\ref{sec:latent_representations}, and we compare the informativeness of the representations using the silhouette score in Sect.~\ref{sec:silhouette_score_results}. In Sect.~\ref{sec:random_forests_macro_recall_results}, we quantify the separation of the classes in the latent representations using the macro recall of random forest classifiers trained on the representations, and we assess the impact of the coefficient variances on the macro recall in Sect.~\ref{sec:impact_of_variances}. Finally, is Sect.~\ref{sec:expected_precision}, we obtain the expected precision of the random forest classifiers on previously unseen data.

\subsection{PCA models of synthetic light curves} \label{sec:pca_models}
\begin{figure}
    \centering
    \includegraphics[width=0.45\textwidth]{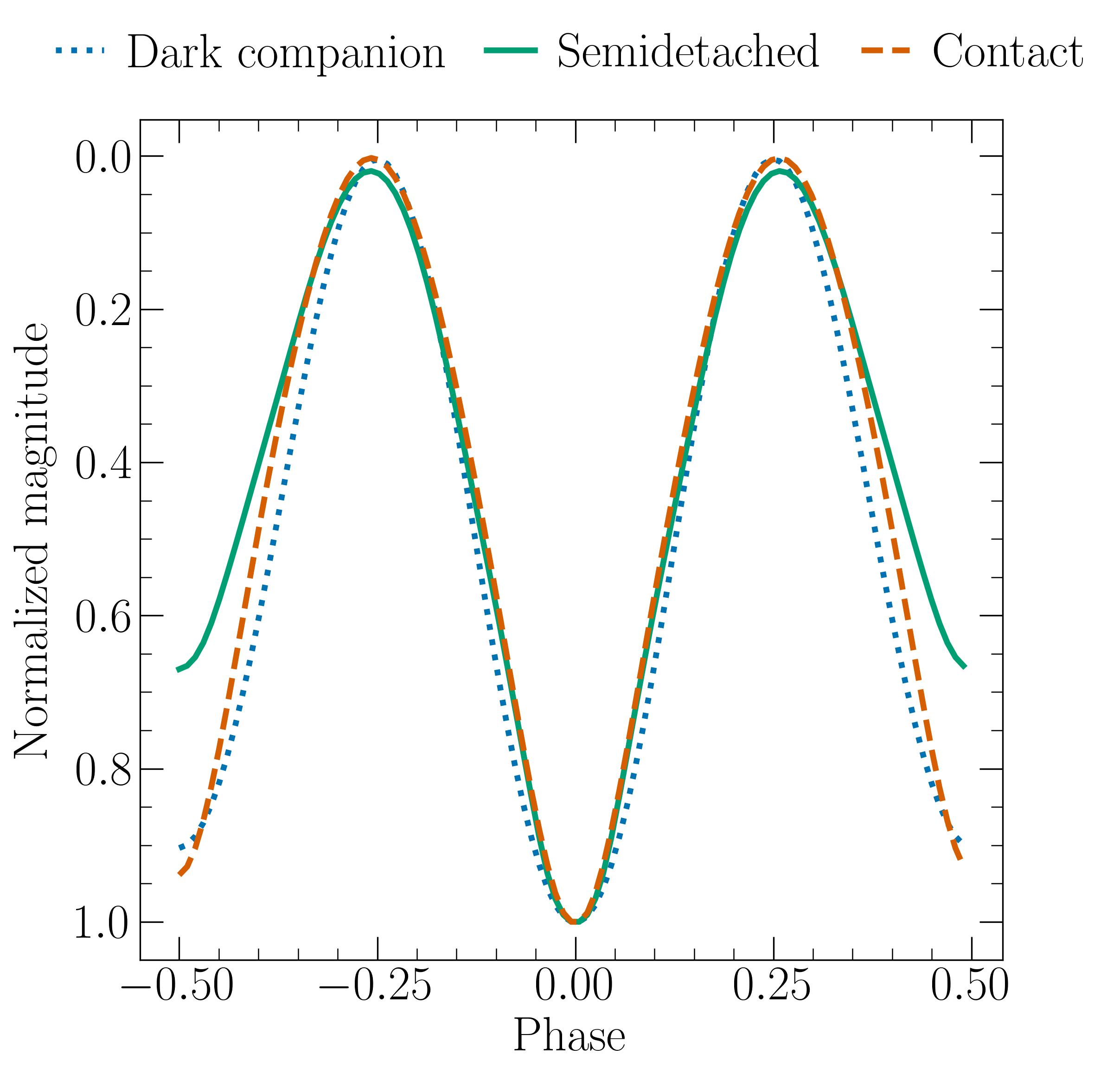}
    \caption{Mean normalized light curves of dark companion, semidetached, and contact binary systems, obtained by taking the average of all normalized light curves of a given class at each phase point of the grid. The mean light curves are subtracted from the light curves of the corresponding class before performing PCA.
    \label{fig:pca_means}}
\end{figure}

\begin{figure*}
    \centering
    \includegraphics[width=0.95\textwidth]{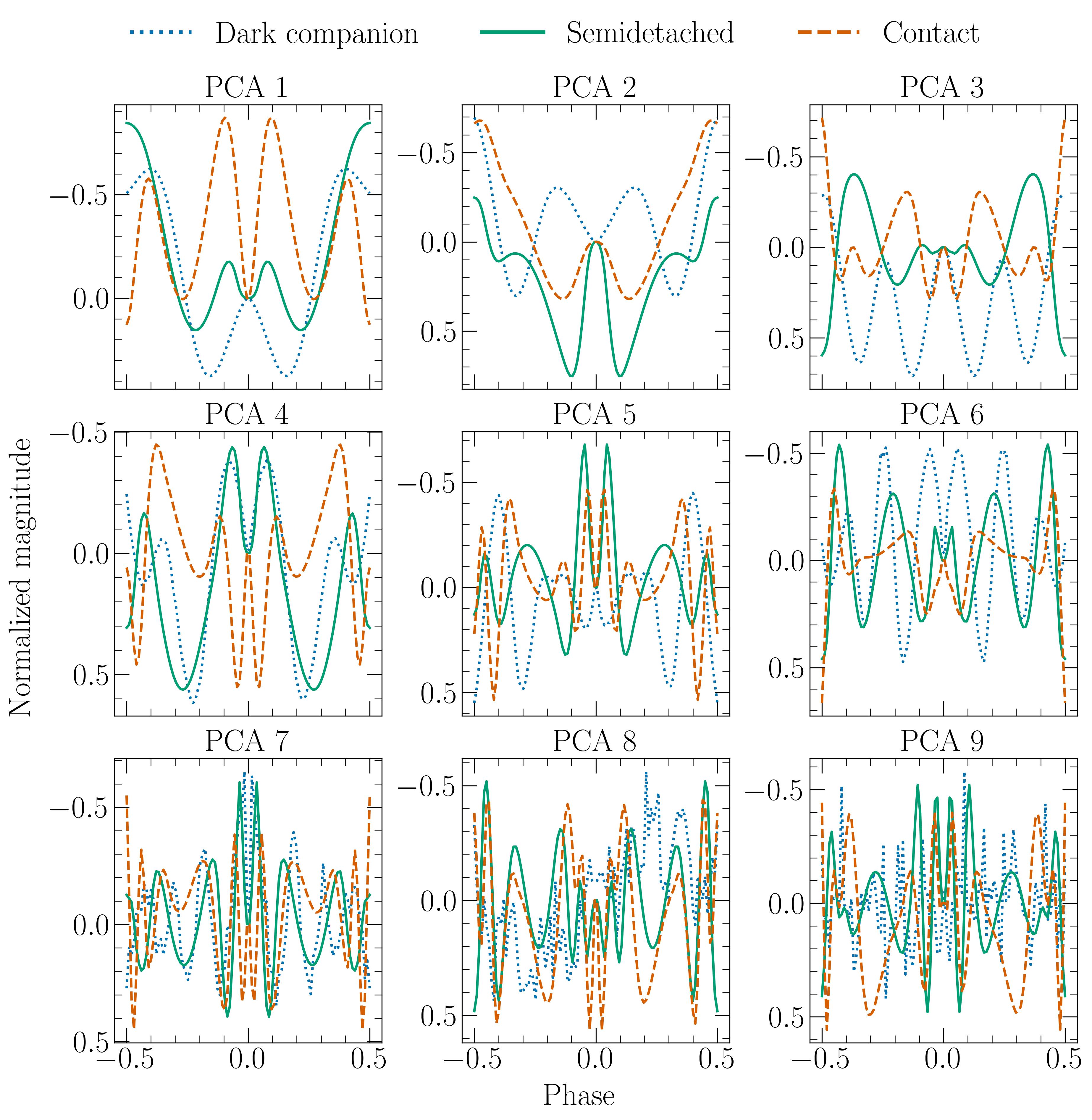}
    \caption{First nine principal components of dark companion, semidetached, and contact binary light curves, ordered by explained variance in descending order. 
    \label{fig:pca_components}}
\end{figure*}

We show the mean normalized light curves of dark companion, semidetached, and contact binary systems in Fig.~\ref{fig:pca_means}. The light curves exhibit a remarkable similarity, particularly between the mean dark companion and contact binary light curves, highlighting the importance of the finer details captured by principal components in distinguishing between the classes. In Fig.~\ref{fig:pca_components}, we show the first nine rescaled principal components of the three binary classes. The principal components have a unit amplitude and are unique up to a sign change. Due to the normalization of the synthetic light curves, all components coincide at phase $0$, corresponding to the primary minimum of the light curves. Compared to the mean light curves, we observe significant differences between the classes already in the first few principal components, with the differences becoming more pronounced as we move away from the primary minimum toward the secondary minimum at phases $-0.5$ and $0.5$. Higher-order principal components are progressively more oscillatory, making it harder to interpret the differences between the classes. Still, we observe that the dark companion components are generally more similar to the contact components than to the semidetached components, revealing increased levels of degeneracy between these two classes.

\begin{figure}
    \centering
    \includegraphics[width=0.45\textwidth]{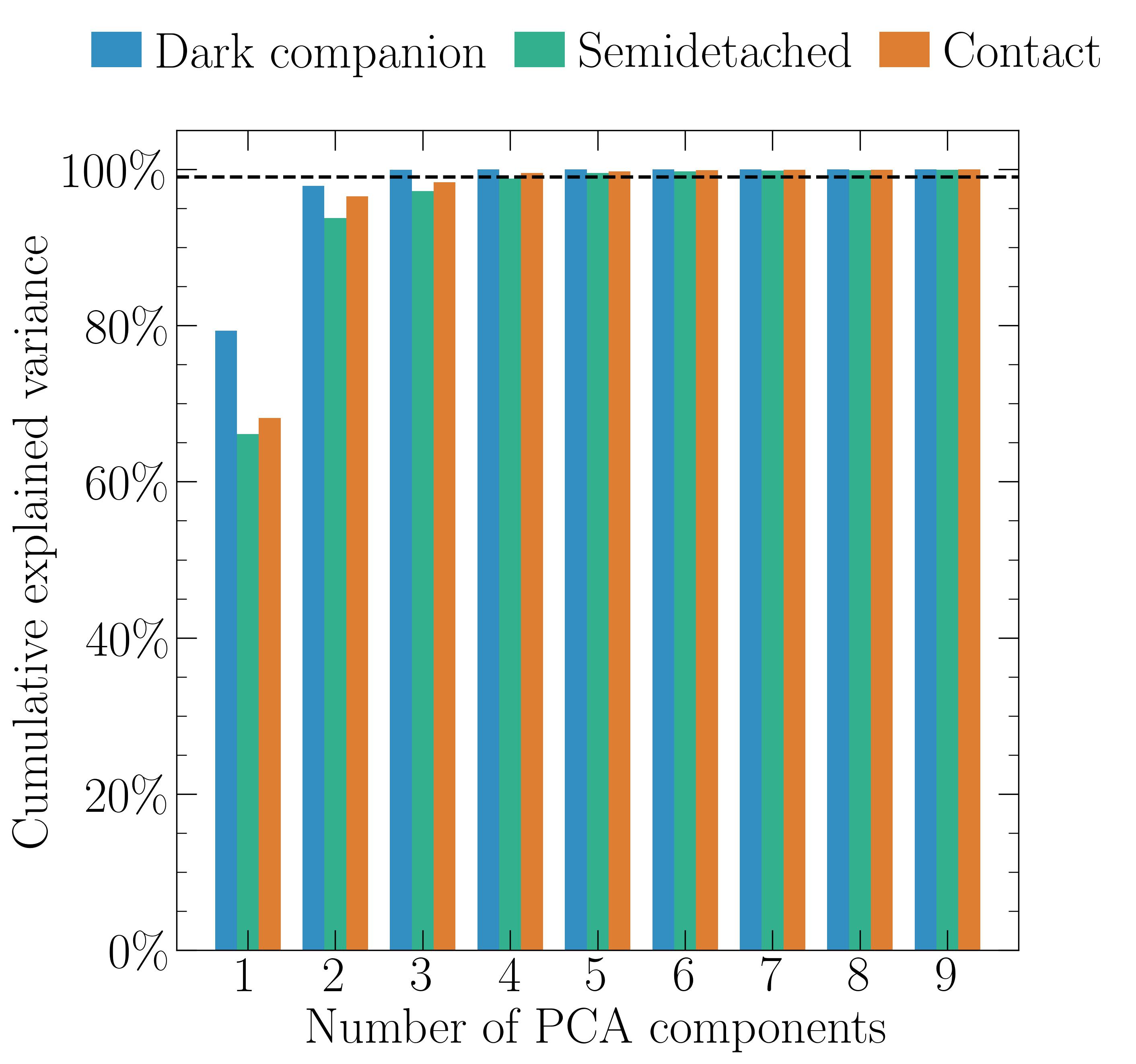}
    \caption{Cumulative explained variance of the principal components of dark companion, semidetached, and contact binary light curves. The dashed line indicates the threshold of $99\%$ explained variance.
    \label{fig:pca_explained_variances}}
\end{figure}

Starting with the seventh component, the dark companion components become increasingly affected by numerical noise, basically becoming pure noise by the ninth component. Consequently, the first six principal components capture virtually all the variance in the dark companion binary light curves. This is not the case for the semidetached and contact binary light curves, which require more components to capture the same level of variance. The disparity in the informativeness of the principal components can be also seen in Fig.~\ref{fig:pca_explained_variances}, which shows the cumulative explained variance as a function of the number of retained PCA components for the three binary classes. The dark companion binary light curves require only three principal components to explain more than $99\%$ of the variance, while the contact and semidetached binary light curves require four and five components, respectively, to cross the $99\%$ threshold.

\subsection{Latent representations} \label{sec:latent_representations}
\begin{figure*}
    \centering
    \includegraphics[width=0.95\textwidth]{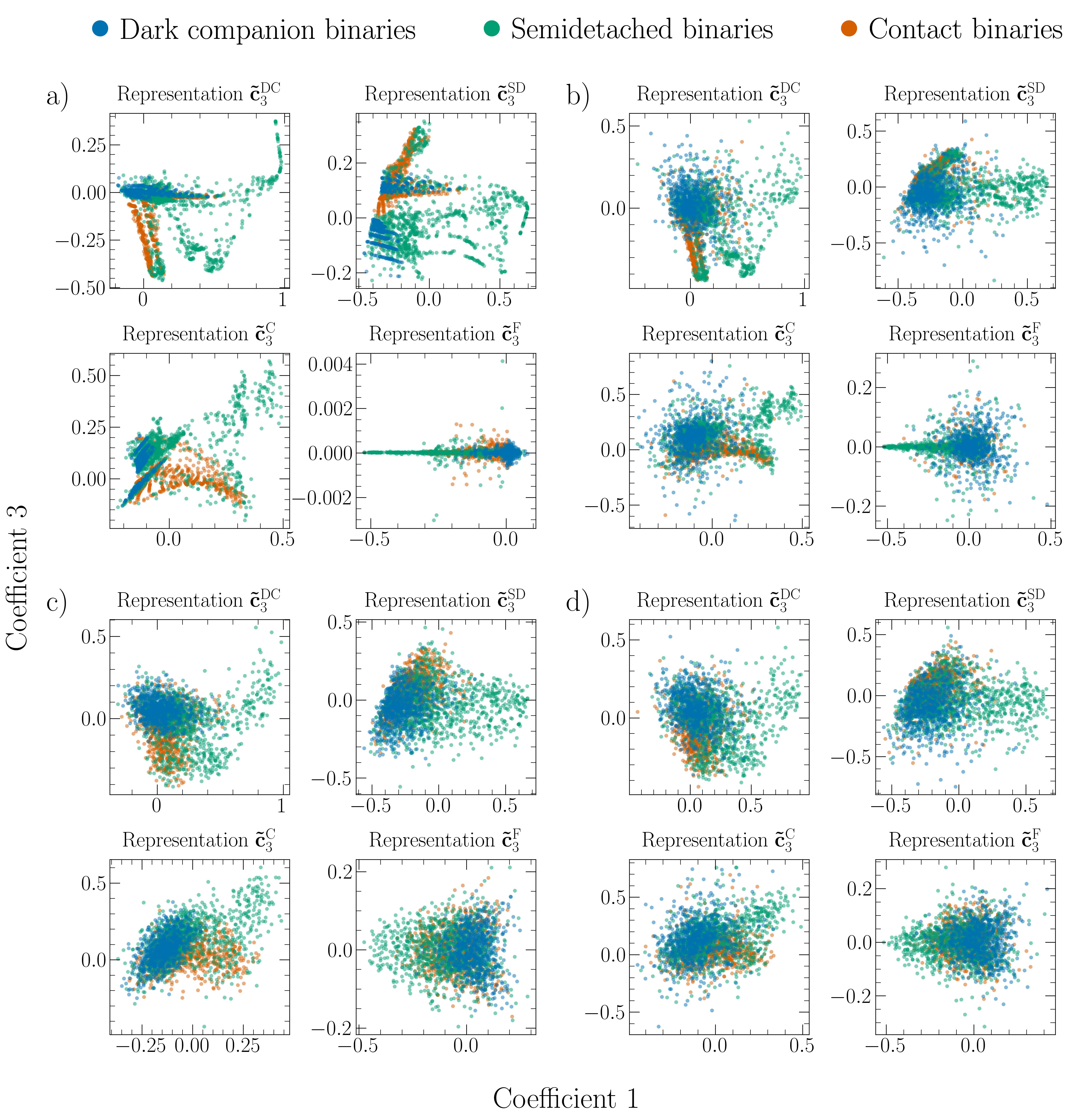}
    \caption{Scatter plots of the first and third coefficients of the representations $\mathbf{\tilde{c}^\text{DC}}$, $\mathbf{\tilde{c}^\text{SD}}$, $\mathbf{\tilde{c}^\text{C}}$, and $\mathbf{\tilde{c}^\text{F}}$ of the dark companion, semidetached, and contact binary light curves in the validation sets of the synthetic samples W0C0 (a), W100C0 (b), W0C10L50 (c), and W100C10L50 (d). We describe the synthetic samples in Sect.~\ref{sec:data}, and we provide the definitions of the representations in Sects.~\ref{sec:pca_representations}--\ref{sec:fourier_representation}.
    \label{fig:scatter_plot_latent_coordinates_1_3_noisy_data_corner_cases}}
\end{figure*}

Utilizing the PCA models and the discretized Fourier basis, we constructed the representations $\mathbf{\tilde{c}}^\text{DC}_{3}$, $\mathbf{\tilde{c}}^\text{SD}_{3}$, $\mathbf{\tilde{c}}^\text{C}_{3}$, and $\mathbf{\tilde{c}}^\text{F}_{3}$ of the validation sets of the four corner cases: samples W0C0, W100C0, W0C10L50, and W100C10L50. In Fig.~\ref{fig:scatter_plot_latent_coordinates_1_3_noisy_data_corner_cases}, we show the scatter plots of the first and third coefficients of the latent representations for these samples. Figure~\ref{fig:scatter_plot_latent_coordinates_1_3_noisy_data_corner_cases}a illustrates the rich structure of the PCA representations of the noiseless sample W0C0, where the dark companion and contact binaries form relatively well-separated clusters, while the semidetached binaries are scattered all over the latent space. In contrast, the Fourier representation is collapsed along the third coefficient and does not show a clear separation between the classes.

Figure~\ref{fig:scatter_plot_latent_coordinates_1_3_noisy_data_corner_cases}b displays the latent representations of the sample W100C0, which was injected with uncorrelated Gaussian noise at $\sigma_\text{WN}$~$=$~$0.01$\,mag. The light curves from this sample exhibit much greater scatter in the latent spaces compared to the noiseless light curves, making it more challenging to differentiate between the classes. Despite this, the PCA representations still retain some of the original structure, especially in the cases of the representations $\mathbf{\tilde{c}}^\text{DC}_{3}$ and $\mathbf{\tilde{c}}^\text{C}_{3}$. In the Fourier representation, the classes are almost completely mixed, with the exception of the semidetached class, which protrudes from the main intermixed cluster. The situation becomes even worse when we introduce correlated noise intended to simulate the effects of surface spots. In Fig.~\ref{fig:scatter_plot_latent_coordinates_1_3_noisy_data_corner_cases}c, we present the latent representations of the sample W0C10L50, which was injected with correlated Gaussian noise at $\sigma_\text{CN}$~$=$~$0.1$ and $l_\text{CN}$~$=$~$0.5$. The Fourier representation of the light curves is practically featureless, with the classes clumped together in a single cluster. The representation $\mathbf{\tilde{c}}^\text{SD}_{3}$ is slightly more informative, but the classes are still thoroughly mixed. We observe the best separation of the classes in the representations $\mathbf{\tilde{c}}^\text{DC}_{3}$ and $\mathbf{\tilde{c}}^\text{C}_{3}$, but the separation is still far from ideal, with most of the structure present in the sample W0C0 lost due to the correlated noise.

Figures~\ref{fig:scatter_plot_latent_coordinates_1_3_noisy_data_corner_cases}b--c demonstrate the independent effects of uncorrelated and correlated noise on the representations of the synthetic light curves. In Fig.~\ref{fig:scatter_plot_latent_coordinates_1_3_noisy_data_corner_cases}d, we show the latent representations of the sample W100C10L50, which incorporates the combined noise from the samples W100C0 and W0C10L50. The cumulative effect of the two types of noise is remarkably similar to the effect of correlated noise alone, with the classes only slightly more mixed in all representations. We conclude that correlated noise, such as the one arising from surface spots, affects the structure of the latent representations more severely than uncorrelated noise, disrupting the patterns present in the absence of noise and effectively mixing the classes.

\begin{figure*}
    \centering
    \includegraphics[width=1\textwidth]{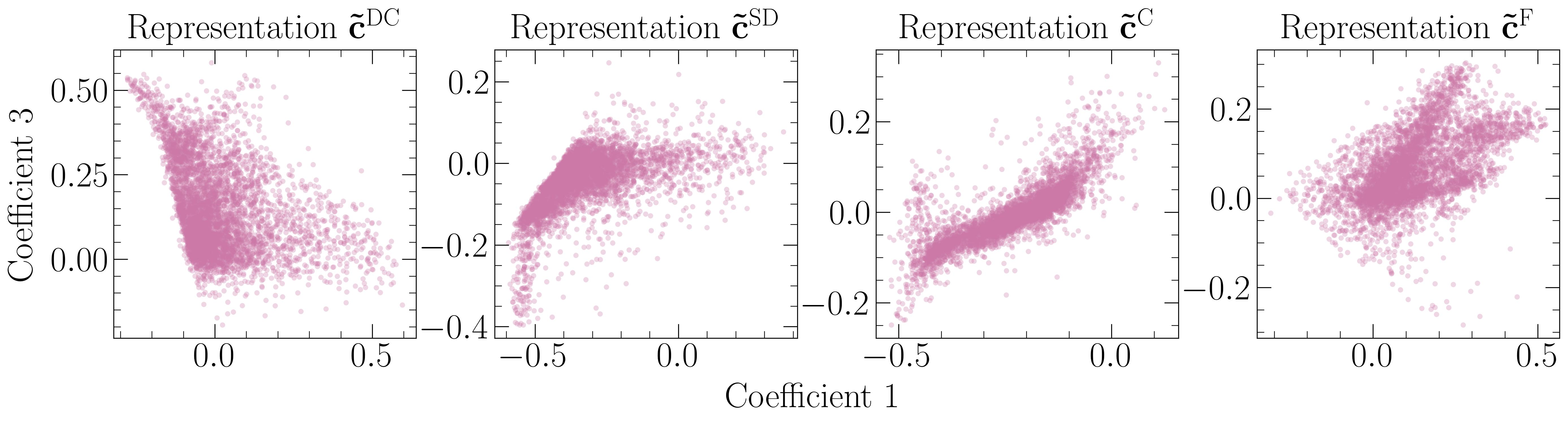}
    \caption{Scatter plots of the first and third coefficients of the representations $\mathbf{\tilde{c}^\text{DC}}$, $\mathbf{\tilde{c}^\text{SD}}$, $\mathbf{\tilde{c}^\text{C}}$, and $\mathbf{\tilde{c}^\text{F}}$ of the TESS ellipsoidal sample from \citet{Green_2023}. We provide the definitions of the representations in Sects.~\ref{sec:pca_representations}--\ref{sec:fourier_representation}.
    \label{fig:scatter_plot_coefficients_1_3_TESS_ellipsoidal_variables}}
\end{figure*}

Although our discussion is based on the visual inspection of the first and third coefficients of the latent representations, the projections to the remaining coefficients yield similar results (Figs.~\ref{fig:scatter_plot_latent_coordinates_1_2_noisy_data_corner_cases}--\ref{fig:scatter_plot_latent_coordinates_2_3_noisy_data_corner_cases}). The only difference is that for some projections, the representation $\mathbf{\tilde{c}}^\text{SD}$ seems to be the most informative, while for others, $\mathbf{\tilde{c}}^\text{DC}$ or $\mathbf{\tilde{c}}^\text{C}$ separate the classes better. In all projections, the Fourier representation $\mathbf{\tilde{c}}^\text{F}$ yields worse visual separation than the most informative PCA representation, demonstrating the superiority of the PCA representations in capturing the latent structure of the synthetic light curves.

To verify that the synthetic samples capture the full range of light curve shapes present in real data, we constructed the PCA and Fourier representations of the TESS sample of ellipsoidal variables from \citet{Green_2023}. The low-dimensional nature of the latent representations makes it possible to visually compare the distributions of real and synthetic light curves in the latent space, providing a simple way to assess the representativeness of the synthetic samples. However, there are some caveats to this approach. First, some light curves in the TESS sample exhibit large gaps in the phase coverage, which can affect their latent representations. To mitigate this and to ensure a fair comparison with the synthetic data, we considered only the light curves with a maximum phase difference between two consecutive observations of less than $0.01$. Second, the TESS sample is not labeled, which means we cannot directly compare the distributions of the classes. In future work, we plan to utilize the methods developed in this work to identify the dark companion binaries in the TESS sample, which will allow for a more direct comparison, but for now, we can only compare the overall shapes of the distributions. Third, we do not know the noise characteristics of the TESS sample, hence we cannot expect any one synthetic sample to perfectly match the distribution of the TESS sample. As a result, it only makes sense to compare the TESS sample with the worst case synthetic scenario -- sample W100C10L50 -- in which the distribution of the light curves in the latent space is most spread out.

In Fig.~\ref{fig:scatter_plot_coefficients_1_3_TESS_ellipsoidal_variables}, we show the scatter plots of the first and third coefficients of the latent representations of the TESS sample. Comparing the plots with Fig.~\ref{fig:scatter_plot_latent_coordinates_1_3_noisy_data_corner_cases}d, we observe that the TESS sample populates roughly the same regions of the latent space as the sample W100C10L50, pointing to a significant overlap with the synthetic data. In addition, the representations of the two samples share some common features, such as the elongated shape of the contact binary representation or the protrusion from the main cluster in the semidetached binary representation. There are also some differences, for example, the contact binary representation of the TESS sample does not extend as far along the first coefficient as the representation of the synthetic sample and the Fourier representation shows a distinct three-pronged structure not present in W100C10L50, which is most likely due to the different noise characteristics of the TESS sample and our synthetic data, but further analysis is needed to confirm this. Still, given the limitations of the comparison, the TESS sample shows a good agreement with the synthetic data. Projections to other combinations of latent coefficients yield similar results (Fig.~\ref{fig:scatter_plot_latent_coordinates_TESS_ellipsoidal_variables}), suggesting that the synthetic samples can be, to a large extent, considered representative of real data.

\subsection{Silhouette scores} \label{sec:silhouette_score_results}
\begin{figure*}
    \centering
    \includegraphics[width=1\textwidth]{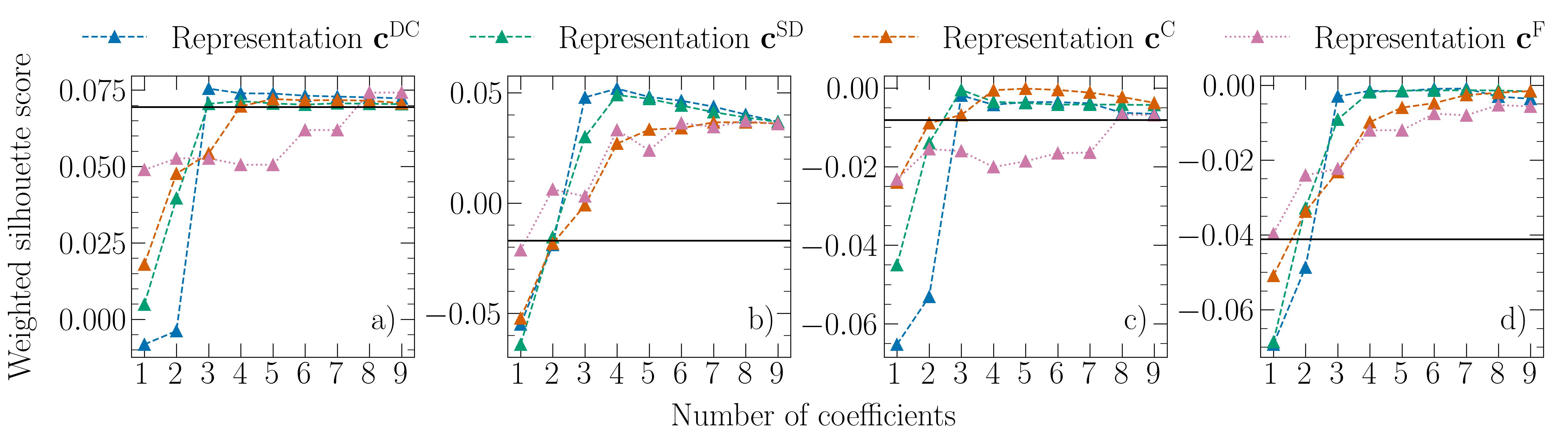}
    \caption{Weighted silhouette scores for the representations $\mathbf{c}^\text{DC}$, $\mathbf{c}^\text{SD}$, $\mathbf{c}^\text{C}$, and $\mathbf{c}^\text{F}$ of the validation sets of the synthetic samples W0C0 (a), W100C0 (b), W0C10L50 (c), and W100C10L50 (d), evaluated as a function of the number of coefficients in the representation. The solid black lines represent the weighted silhouette scores calculated for the full normalized light curves in the validation sets. We describe the synthetic samples in Sect.~\ref{sec:data}, and we define the representations in Sects.~\ref{sec:pca_representations}--\ref{sec:fourier_representation}.
    \label{fig:silhouette_scores_noisy_data_corner_cases}}
\end{figure*}

While visual inspection of the latent representations can provide qualitative insight into the separation of the dark companion, semidetached, and contact binary light curves, the silhouette score allows for a more quantitative and systematic approach to assessing the separation of the classes in the different representations. In Fig.~\ref{fig:silhouette_scores_noisy_data_corner_cases}, we show the silhouette scores for the unit representations $\mathbf{c}^\text{DC}_{n}$, $\mathbf{c}^\text{SD}_{n}$, $\mathbf{c}^\text{C}_{n}$, and $\mathbf{c}^\text{F}_{n}$ of the validation sets of the corner cases, evaluated as a function of the number of coefficients in the representation $n$~$=$~$1\text{--}9$. The solid black lines represent the silhouette scores for the full representations of the samples, which vary from approximately $0.07$ in the case of the sample W0C0 (Fig.~\ref{fig:silhouette_scores_noisy_data_corner_cases}a) to about $-0.04$ for the sample W100C10L50 (Fig.~\ref{fig:silhouette_scores_noisy_data_corner_cases}d). Taken at face value, the silhouette scores of the full representations are low, indicating that the classes are not well separated in the original high-dimensional space. However, as we discuss in Sect.~\ref{sec:silhouette_score_methods} and Appendix~\ref{app:silhouette_score}, the absolute value of the silhouette score is not important, only the difference between the silhouette scores of different representations is informative. Thus, we use the silhouette score of the full representation as a benchmark against which we compare the silhouette scores of the latent representations to see whether projection to a lower-dimensional space can improve the separation of the classes.

Due to the orthogonal character of the unit PCA and Fourier representations, the silhouette scores of the representations converge to the silhouette score of the full representation as the number of coefficients $n$ goes to $N_\text{grid}$. The reason is that the distance in the definition of the silhouette score is calculated using the dot product of the difference vector with itself, which is preserved under orthogonal transformations. In the case of the synthetic samples W0C0 and W0C10L50 (Figs.~\ref{fig:silhouette_scores_noisy_data_corner_cases}a and c), the silhouette scores of the PCA representations quickly plateau slightly above the benchmark limit value. The situation is different for the samples W100C0 and W100C10L50 (Fig.~\ref{fig:silhouette_scores_noisy_data_corner_cases}b and d), where the silhouette scores of all latent representations peak high above the benchmark score before reaching the limit value at $n$~$=$~$N_\text{grid}$. In all four corner cases, the silhouette score of the representation $\mathbf{c}^\text{F}$ follows similar trends to the silhouette scores of the PCA representations, but it requires more coefficients to reach the same levels of class separation, with the exception of the first one to two coefficients, which seem to be more informative than the PCA coefficients.

The general trends in the silhouette scores of the latent representations under different noise conditions can be explained by the properties of the injected noise or the lack thereof. In the absence of noise (Fig.~\ref{fig:silhouette_scores_noisy_data_corner_cases}a), the PCA representations require only a few coefficients to almost perfectly reconstruct the light curves, leaving only negligible unexplained variance to be captured by higher-order coefficients. Consequently, the higher-order coefficients are close to zero and do not significantly contribute to the silhouette score, which explains the quick plateauing of the silhouette scores. In other words, the first three to four coefficients, depending on the PCA representation, capture effectively all the information that is present in the full light curves and account for most of the separation between the classes. Conversely, the representation $\mathbf{c}^{F}$ requires at least eight coefficients to reach the plateau, pointing to the poor alignment of the Fourier basis with the data.

The power spectrum of a light curve injected with noise is the sum of the power spectrum of the signal, the power spectrum of the noise, and an additional term arising from the interaction of the signal and the noise. While the power spectrum of the signal is skewed toward low frequencies, the power spectrum of uncorrelated noise is flat, which means that, in relative terms, low-order coefficients of the PCA representations are less affected by the noise than higher-order coefficients. Given that most of the signal is contained in the first three to five coefficients (Fig.~\ref{fig:pca_explained_variances}), we are able to extract useful information from light curves even in the presence of strong uncorrelated noise. This can be seen in Fig.~\ref{fig:silhouette_scores_noisy_data_corner_cases}b, where the silhouette scores of the representations $\mathbf{c}^\text{DC}$ and $\mathbf{c}^\text{SD}$ peak at $n$~$=$~$4$ and then start to decrease as we keep adding coefficients that are increasingly more affected by the noise. Compared to the case with no noise, the contribution of the higher-order coefficients to the silhouette score is not negligible, because they capture the high-frequency noise that is absent in the former case. Given that the high-frequency noise affects all classes equally, the classes become gradually more mixed together as we increase $n$, resulting in the low benchmark silhouette score of the full representation. The silhouette scores of the representations $\mathbf{c}^\text{C}$ and $\mathbf{c}^\text{F}$ follow a similar trend, but they require more coefficients to reach the same levels of class separation as $\mathbf{c}^\text{DC}$ and $\mathbf{c}^\text{SD}$, if at all, revealing a suboptimal alignment of their bases with the data.

The situation is different when we inject the light curves with strong correlated noise, where the power spectra of both the signal and the noise are skewed toward low frequencies (Fig.~\ref{fig:silhouette_scores_noisy_data_corner_cases}c). In this case, the noise effectively masks the signal in the low-order coefficients of the PCA representations, preventing us from recovering the original signal. Not being able to distinguish between the signal and the noise, the representations treat the noise as part of the signal, resulting in a scenario analogous to the noiseless case. The first few PCA coefficients are enough to capture virtually all variance in the light curves, pushing higher-order coefficients to zero. Consequently, the silhouette scores of the PCA representations settle slightly above the benchmark score, whose value is significantly lower than in the noiseless case. Conversely, the silhouette score of the Fourier representation $\mathbf{c}^\text{F}$ approaches the benchmark from below and requires at least eight coefficients  to reach the plateau.

The combined effect of strong correlated and uncorrelated noise is to decrease the separation between the classes even further, pushing the benchmark score to decidedly negative values. (Fig.~\ref{fig:silhouette_scores_noisy_data_corner_cases}d). Despite the low benchmark score, the silhouette scores of the latent representations are comparable to the values obtained for the sample W0C10L50, revealing that in the presence of both types of noise i) the latent representations still manage to disentangle low-frequency signal from high-frequency noise and ii) the level of the correlated noise is the decisive factor in determining the structure of the latent representations and the separation of the classes. The latter is consistent with our visual inspection of the corner cases in the previous section, where the latent representations of the samples W0C10L50 and W100C10L50 exhibited similar features.

Based on our analysis of the silhouette scores in the four corner cases, the three PCA representations perform comparably. Although the representations $\mathbf{c}^\text{DC}$ and $\mathbf{c}^\text{SD}$ outperform $\mathbf{c}^\text{C}$ on the sample W100C0, the differences between their maximum silhouette scores are marginal in all other corner cases, pointing to the equivalence of the three representations in terms of class separation. Regarding the Fourier representation $\mathbf{c}^\text{F}$, for $n$~$>$~$2$ it generally achieves worse silhouette scores than the best performing PCA representation for the same number of coefficients, with the exception of the sample W0C0, where it performs comparably at $n$~$=$~$8\text{--}9$. Still, $\mathbf{c}^\text{F}$ does not achieve the maximum silhouette score in any of the corner cases, which is consistent with our visual inspection of the latent representations in the previous section, where the Fourier representation consistently yielded worse class separation than the most discriminative PCA representation.

\subsection{Macro recalls and random forest hyperparameters} \label{sec:random_forests_macro_recall_results}
\begin{figure*}[ht!]
    \centering
    \includegraphics[width=1\textwidth]{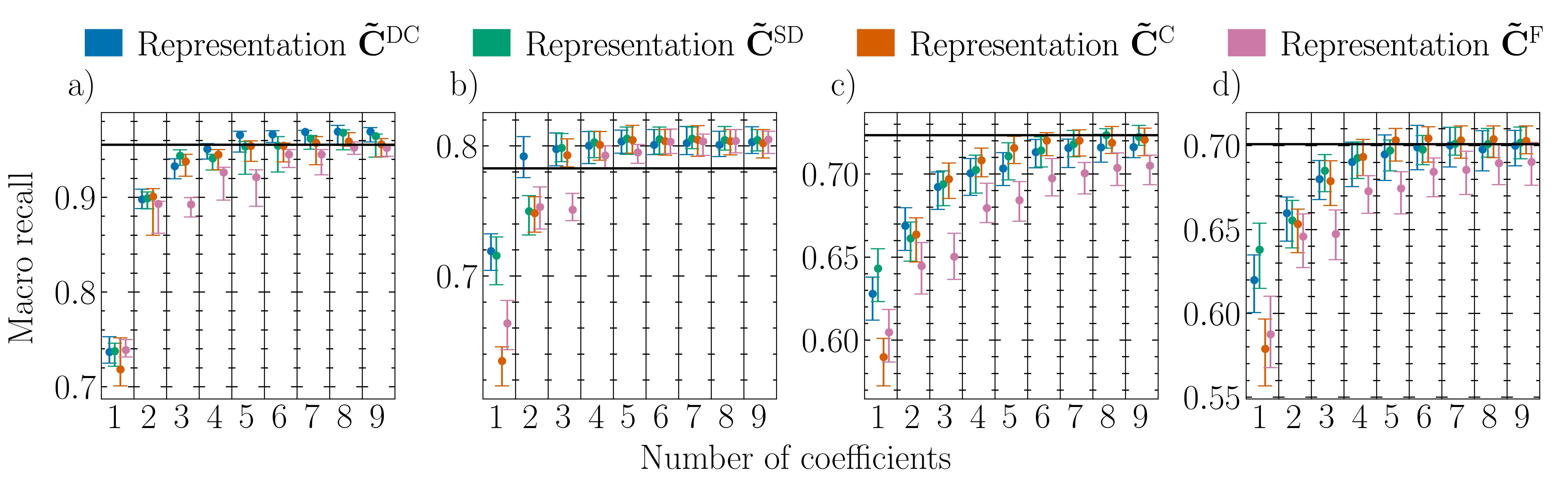}
    \caption{
    Validation macro recalls of the random forest classifiers trained on the representations $\mathbf{\tilde{C}^\text{DC}}$, $\mathbf{\tilde{C}^\text{SD}}$, $\mathbf{\tilde{C}^\text{C}}$, and $\mathbf{\tilde{C}^\text{F}}$ of the synthetic samples W0C0 (a), W100C0 (b), W0C10L50 (c), and W100C10L50 (d), evaluated as a function of the number of coefficients in the representation. The errorbars show the minimum and the maximum validation macro recalls for a given number of coefficients. The solid black lines represent the best validation macro recalls achieved by the classifiers trained on the extended full representations of the samples. We describe the synthetic samples in Sect.~\ref{sec:data}, and we provide the definitions of the representations in Sects.~\ref{sec:pca_representations}--\ref{sec:fourier_representation}.
    \label{fig:macro_recalls_corner_cases_only_coeffs}}
\end{figure*}

The silhouette score is useful for comparing the separation of the dark companion, semidetached, and contact binary classes across different latent representations, but it does not provide an absolute measure of class separation. To address this, we trained random forest classifiers on the extended representations $\mathbf{\tilde{C}}^\text{DC}_{n}$, $\mathbf{\tilde{C}}^\text{SD}_{n}$, $\mathbf{\tilde{C}}^\text{C}_{n}$, and $\mathbf{\tilde{C}}^\text{F}_{n}$ of the training sets of all synthetic samples (Table~\ref{tab:synthetic_samples}) for $n$~$=$~$1\text{--}9$, and we conducted a basic hyperparameter search on the validation sets of the samples to find the configurations that yield the best macro recalls. In Fig.~\ref{fig:macro_recalls_corner_cases_only_coeffs}, we present the obtained validation macro recalls of the random forest classifiers as a function of $n$ for the four corner cases. The errorbars show the minimum and the maximum validation macro recalls achieved by the classifiers during the hyperparameter tuning. The solid black lines represent the best macro recalls achieved by the classifiers trained on the extended full representations. We also trained random forest classifiers on the one-dimensional representations of the samples, but we do not show the results in the plots, because they make them less readable. The best validation macro recalls achieved by the classifiers trained on the one-dimensional representations are: $R^\text{V}_\text{M}$~$=$~$0.49$ for the noiseless sample W0C0 (Fig.~\ref{fig:macro_recalls_corner_cases_only_coeffs}a), $R^\text{V}_\text{M}$~$=$~$0.52$ for the sample W100C0 (Fig.~\ref{fig:macro_recalls_corner_cases_only_coeffs}b) and $R^\text{V}_\text{M}$~$=$~$0.53$ for the remaining corner cases (Fig.~\ref{fig:macro_recalls_corner_cases_only_coeffs}c--d).

In all corner cases, the macro recalls of the PCA representations quickly reach a saturation point somewhere between $n$~$=$~$3$ and $6$, and then they level off with a slight increase or decrease. We observed similar trends in the silhouette scores of the PCA representations, but the saturation points of the macro recalls are generally shifted toward higher $n$ compared to the peaks and plateaus in the silhouette scores. The representations of the samples W0C0 and W0C10L50 exhibit the most pronounced shifts, with the macro recall saturation points at $n$~$=$~$5\text{--}6$ and the silhouette score plateaus at $n$~$=$~$3\text{--}4$. Another difference between the silhouette scores and the macro recalls of random forest classifiers is that the latter are not as sensitive to the presence of uninformative features. This can be seen by comparing the trends in the silhouette scores and the macro recalls for the PCA representations of the samples W100C0 and W100C10L50. The silhouette scores of the PCA representations peak high above the benchmark silhouette score of the full representation and then start to decrease, eventually reaching the limit value at $n$~$=$~$N_\text{grid}$ (Figs.~\ref{fig:silhouette_scores_noisy_data_corner_cases}b and d, the decrease to the limit value not shown), while the macro recalls quickly reach a plateau located slightly above the limit value of the best macro recall achieved for the extended full representation, allowing only for a marginal decrease with additional coefficients (Figs.~\ref{fig:macro_recalls_corner_cases_only_coeffs}b and d).

Both the shift of the saturation points toward higher $n$ and the generally nondecreasing trends in the macro recalls of the PCA representations with increasing $n$ can be explained by the nonlinear nature of the random forest classifier and its robustness to noise and overfitting. The nonlinear nature allows the classifier to learn complex decision boundaries that are not necessarily convex nor connected, making it possible to extract useful information even from coefficients that decrease the silhouette score of the representation. The robustness to noise and overfitting means that even if we include an uninformative feature, the classifier can learn to ignore it and focus on the informative features, leaving the macro recall essentially unchanged. These arguments also extend to the macro recall of the representation $\mathbf{\tilde{C}}^\text{F}$, which follows the same nondecreasing trend with increasing $n$, but consistently requires more coefficients to reach the same levels as the macro recalls of the PCA representations. The only exception is the sample W100C0, where $\mathbf{\tilde{C}}^\text{F}$ performs comparably to the PCA representations. Unlike the PCA representations, $\mathbf{\tilde{C}}^\text{F}$ does not show a significant shift in the macro recall trends toward higher $n$ compared to the trends in the silhouette score, indicating that the silhouette score is relatively well-aligned with how the data is organized in the Fourier latent space.

In Table~\ref{tab:results}, we present the optimal representations and random forest hyperparameters that yielded the best validation macro recalls for each synthetic sample. We also present the class and macro recalls of the best performing classifiers evaluated on the validation and test sets of the synthetic samples, demonstrating the generalization of the recalls to previously unseen data. The optimal number of trees in the forest is \texttt{n\_estimators}~$=$~\texttt{500} in the majority of cases, but for some samples, \texttt{n\_estimators}~$=$~\texttt{100} yields better results. We do not observe a clear pattern between the optimal value of \texttt{n\_estimators} and the level or type of noise, indicating that the random forest classifier is robust to the choice of this hyperparameter in the context of the synthetic samples. The optimal minimum number of samples at a leaf node alternates between \texttt{min\_samples\_leaf}~$=$~\texttt{1} and \texttt{10}, but in the presence of moderately strong to strong uncorrelated noise ($\sigma_\text{WN}$~$\gtrsim$~$10^{-3}$\,mag) and/or correlated noise ($\sigma_\text{CN}$~$\gtrsim$~$0.05$ and $l_\text{CN}$~$\leq$~$0.5$), the optimal value is preferentially \texttt{min\_samples\_leaf}~$=$~\texttt{10}, reducing the risk of overfitting. In the majority of noise conditions, the optimal method for selecting the number of features at each split is \texttt{max\_features}~$=$~\texttt{sqrt}, but in the limit of strong uncorrelated and correlated noise, \texttt{max\_features}~$=$~\texttt{log2} is also a valid choice. Overall, the setup with \texttt{n\_estimators}~$=$~\texttt{500}, \texttt{min\_samples\_leaf}~$=$~\texttt{10}, and \texttt{max\_features}~$=$~\texttt{sqrt} seems to be the most robust, yielding the best macro recalls for $12$ out of $40$ synthetic samples across a wide range of noise conditions.

\begin{figure*}[ht!]
    \centering
    \includegraphics[width=0.95\textwidth]{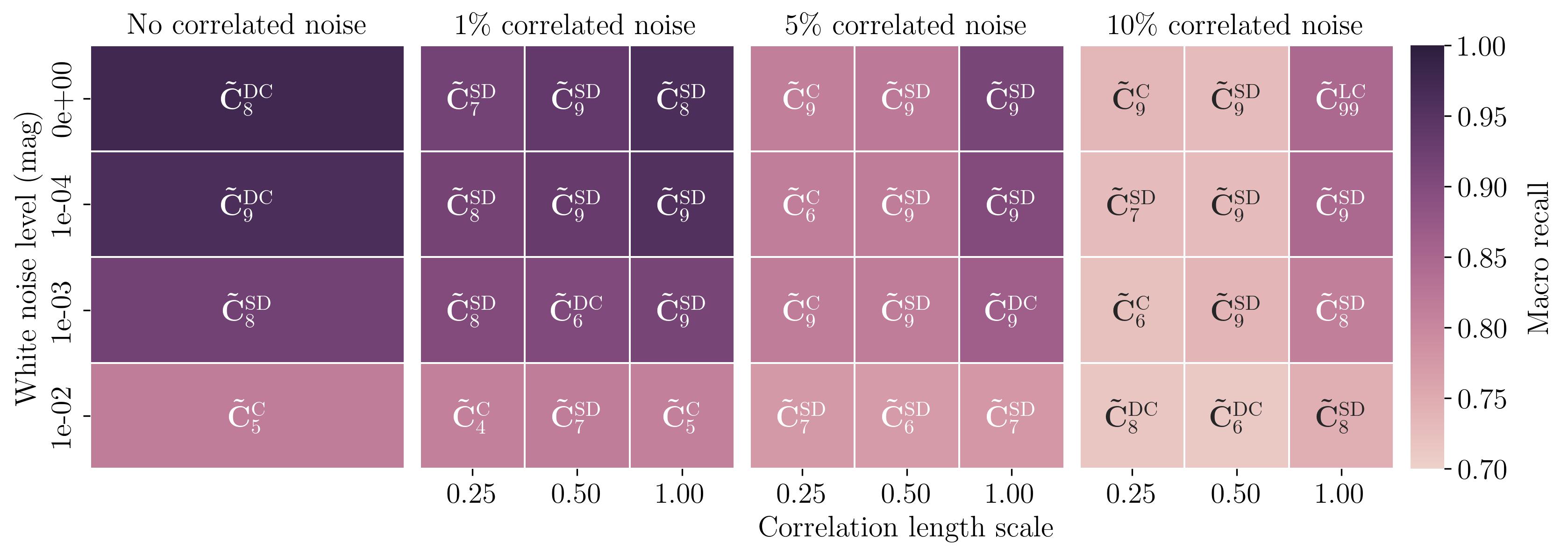}
    \caption{
    Latent representations that yielded the best macro recalls on the validation sets of the synthetic samples. We describe the synthetic samples in Sect.~\ref{sec:data}, and we provide the definitions of the representations in Sects.~\ref{sec:pca_representations}--\ref{sec:fourier_representation}. The symbol $\mathbf{\tilde{C}}^\text{LC}_{99}$ denotes the extended full representation.
    \label{fig:representations_and_dimensions_at_best_recalls}}
\end{figure*}

To visualize the optimal representations of the synthetic samples, we present Fig.~\ref{fig:representations_and_dimensions_at_best_recalls}. The figure shows the representations that yielded the best validation macro recalls for each synthetic sample, along with the corresponding dimensions at which these recalls were attained. The samples are \mbox{color-coded} according to the highest macro recall achieved on their validation sets. In most cases ($25$ out of $40$), the best macro recalls are achieved for the representation $\mathbf{\tilde{C}}^\text{SD}$. The superior performance of $\mathbf{\tilde{C}}^\text{SD}$ is most likely due to the increased variance of semidetached binary light curves compared to dark companion and contact binary light curves (Fig.~\ref{fig:pca_explained_variances}), resulting in more informative higher-order principal components that are less affected by numerical noise and are able to better capture high-frequency information in the light curves. However, as can be seen in Fig.~\ref{fig:macro_recalls_corner_cases_only_coeffs}, the difference between the best macro recalls achieved by the representations is only marginal, rendering the choice of the PCA representation non-critical. Irrespective of the noise conditions, the best performing PCA representation consistently outperforms $\mathbf{\tilde{C}}^\text{F}$ as well as the one-dimensional representation and the extended full representation, except for the case of the sample W0C10L100, where the extended full representation yields the best macro recall.

Regarding the optimal number of coefficients in the latent representations, we observe that the best macro recalls are generally achieved for $n$~$=$~$7\text{--}9$, but as few as four coefficients are enough in the case of the sample W100C1L25. The relatively high numbers of coefficients required to achieve the best macro recalls can be explained by the nonlinear nature of the random forest classifier and its robustness to overfitting. These properties ensure that the classifier generally does not perform significantly worse when trained on a representation with more coefficients, even if the increase in the macro recall is only marginal and the classifier would perform comparably well with fewer coefficients. It is possible that in some cases, the optimal number of coefficients is even greater than nine, but we did not explore this possibility further. Even if that was the case, the marginal increase in the macro recall beyond the saturation points (Fig.~\ref{fig:macro_recalls_corner_cases_only_coeffs}) makes the analysis largely redundant. Consequently, the random forest classifier is robust to the choice of the number of coefficients in the latent representations, provided we are in the saturated macro recall regime. As a rule of thumb, seven to nine coefficients should be sufficient to capture all the relevant information while avoiding overfitting, but in the presence of strong noise, fewer coefficients may be more appropriate.

In Fig.~\ref{fig:validation_test_macro_recalls}, we show the best macro recalls achieved by the random forest classifiers on the validation sets of the synthetic samples (top panel) and the macro recalls of the best performing classifiers on the test sets (bottom panel). By evaluating the macro recalls on the test sets, we obtain a more realistic estimate of the class overlap in the latent representations. We observe that the test macro recalls are generally lower than the validation macro recalls, but the difference is relatively low, with a maximum decrease of $0.06$ in absolute terms, demonstrating reliable generalization to previously unseen data. In the presence of purely uncorrelated noise, the test macro recalls vary from $R^\text{T}_\text{M}$~$=$~$0.96$ to $0.78$, while for purely correlated noise, the macro recalls range from $R^\text{T}_\text{M}$~$=$~$0.97$ to $0.69$, depending on $\sigma_\text{CN}$ and $l_\text{CN}$. The general trend is that for fixed $\sigma_\text{WN}$ and $\sigma_\text{CN}$, the test macro recalls decrease with decreasing $l_\text{CN}$, revealing that the effect of correlated noise is more adverse for shorter correlation length scales. Overall, the test macro recalls of the best performing random forest classifiers range from $R^\text{T}_\text{M}$~$=$~$0.97$ (noiseless sample W0C0) down to $R^\text{T}_\text{M}$~$=$~$0.67$ (sample W100C10L25), indicating a moderately low overlap of the classes in the latent space even in the presence of high levels of uncorrelated and correlated noise.

\subsection{Impact of variances on macro recalls} \label{sec:impact_of_variances}
\begin{figure*}[ht!]
    \centering
    \includegraphics[width=1\textwidth]{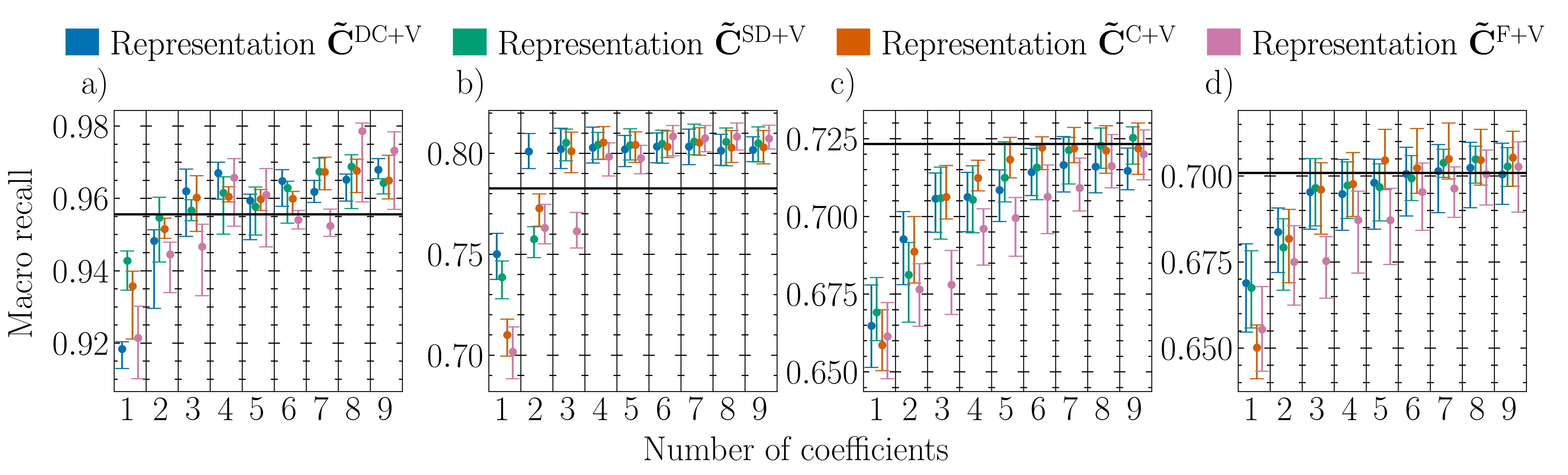}
    \caption{
    Validation macro recalls of the random forest classifiers trained on the augmented representations $\mathbf{\tilde{C}^\text{DC+V}}$, $\mathbf{\tilde{C}^\text{SD+V}}$, $\mathbf{\tilde{C}^\text{C+V}}$, and $\mathbf{\tilde{C}^\text{F+V}}$ of the synthetic samples W0C0 (a), W100C0 (b), W0C10L50 (c), and W100C10L50 (d), evaluated as a function of the number of coefficients in the representation. The errorbars show the minimum and the maximum validation macro recalls for a given number of coefficients. The solid black lines represent the best validation macro recalls achieved by the classifiers trained on extended full representations of the samples. We describe the synthetic samples in Sect.~\ref{sec:data} and provide details about the augmented representations in Sect.~\ref{sec:random_forests_macro_recall_methods}.
    \label{fig:macro_recalls_corner_cases_augmented_representations}}
\end{figure*}

To assess the additional information contained in the uncertainties of the coefficients of the latent representations, we repeated the analysis, including the optimization of the hyperparameters of the random forest classifiers, on the extended latent representations augmented with the variances obtained from the least squares fits of the photometric amplitude and the coefficients. In Fig.~\ref{fig:macro_recalls_corner_cases_augmented_representations}, we show the obtained validation macro recalls as a function of $n$ for the four corner cases. The errorbars and the solid black lines have the same meaning as in Fig.~\ref{fig:macro_recalls_corner_cases_only_coeffs}. For readibility reasons, we do not show the results for the one-dimensional representations. The best validation macro recalls achieved by the classifiers trained on the one-dimensional representations augmented with variances are: $R^\text{V}_\text{M}$~$=$~$0.63$ for the sample W0C0 (Fig.~\ref{fig:macro_recalls_corner_cases_augmented_representations}a), $R^\text{V}_\text{M}$~$=$~$0.60$ for the samples W100C0 and W0C10L50 (Fig.~\ref{fig:macro_recalls_corner_cases_augmented_representations}b--c), and $R^\text{V}_\text{M}$~$=$~$0.59$ for the sample W100C10L50 (Fig.~\ref{fig:macro_recalls_corner_cases_augmented_representations}d). This represents an absolute increase of $0.06$ to $0.14$ in the macro recalls of the one-dimensional representations compared to the unaugmented case. We observe a similar shift to higher macro recalls for all augmented latent representations across all synthetic samples, but the shift becomes less pronounced with increasing $n$. The increase in the macro recalls is most prominent for the noiseless sample W0C0, where the performance of the PCA representations is considerably improved for $n$~$=$~$1\text{--}4$. In the presence of noise, the macro recalls of low-dimensional latent representations are still higher when augmented with variances, but the improvement practically vanishes beyond the saturation points of the unaugmented representations. Conversely, the macro recalls of the Fourier representation are positively affected by the inclusion of variances up to $n$~$=$~$9$, even surpassing the macro recalls of the PCA representations on several occasions.

We present the optimal augmented representations and hyperparameter setups for each synthetic sample in Table~\ref{tab:results_augmented_representations}. The optimal setups are remarkably similar to the setups obtained for the unaugmented representations, with the most significant difference being the occasional preference of \texttt{max\_features}~$=$~\texttt{None} instead of \texttt{max\_features}~$=$~\texttt{log2} in the limit of strong noise. Other than that, the optimal setups remain largely unchanged, with \texttt{n\_estimators}~$=$~\texttt{500}, \texttt{min\_samples\_leaf}~$=$~\texttt{10}, and \texttt{max\_features}~$=$~\texttt{sqrt} leading to the best results for $12$ out of $40$ synthetic samples. Similarly, the semidetached binary representation remains the most informative representation, achieving the best macro recall in $19$ out of $40$ cases (Fig.~\ref{fig:augmented_representations_and_dimensions_at_best_recalls}). However, it is the Fourier representation that benefits the most from the inclusion of variances, yielding the best macro recall for $13$ samples -- a significant improvement over the unaugmented case, where it did not manage to outperform the PCA representations for any of the samples.

Similar to the representations without variances, the augmented representations achieve the best results for $n$~$=$~$7\text{--}9$. This is not surprising, given the limited positive impact of the variances on the macro recalls beyond the saturation points of the unaugmented representations at $n$~$=$~$3\text{--}6$ (Figs.~\ref{fig:macro_recalls_corner_cases_only_coeffs} and \ref{fig:macro_recalls_corner_cases_augmented_representations}). Consequently, the best validation macro recalls achieved by the random forest classifiers and the test macro recalls of the best performing classifiers are almost identical whether we augment the representations with variances or not (Figs.~\ref{fig:validation_test_macro_recalls} and \ref{fig:validation_test_macro_recalls_augmented_representations}). Even in the noiseless case, where the impact of the variances is the most pronounced, the improvement is only marginal. In some cases, the best macro recall even decreases when we include the variances (e.g., sample W100C0).

We conclude that while the inclusion of variances can significantly improve the macro recall of random forest classifiers trained on low-dimensional latent representations ($n$~$\lesssim$~$3$), the positive effect diminishes with increasing dimension of the representation, becoming negligible once we cross the macro recall saturation points of the unaugmented representations at $n$~$=$~$3\text{--}6$. Consequently, augmenting the latent representations with variances is generally not necessary, provided the dimension of the representation is sufficiently high to capture all the relevant information. To achieve the best results, we recommend training the classifier on latent representations with $n$~$=$~$7\text{--}9$.

\subsection{Expected precision of random forest classifiers} \label{sec:expected_precision}
The relatively high macro recalls of the random forest classifiers on the validation and test sets of the samples of dark companion, semidetached, and contact binary light curves must be interpreted in the context of the synthetic data, which we deliberately balanced to avoid bias toward a particular class. Consequently, the macro recalls should be understood as a measure of the mean non-overlap of the classes in the latent space rather than the expected accuracy of the classifier on real data. Despite that, we can still use the class recalls obtained for the test sets of the synthetic samples (Table~\ref{tab:results}) to estimate the expected precision of the classifier on previously unseen data, provided we know the relative frequencies of the classes in the sample. Assuming the worst case scenario, where all misclassifications fall into the dark companion class, we can estimate the expected precision $P_\text{DC}$ of the classifier for the dark companion class as
\begin{ceqn}
\begin{equation}
    P_\text{DC} = \frac{f_\text{DC}R^\text{T}_\text{DC}}{f_\text{DC}R^\text{T}_\text{DC}+f_\text{SD}(1-R^\text{T}_\text{SD})+f_\text{C}(1-R^\text{T}_\text{C})},
    \label{eq:expected_precision}
\end{equation}
\end{ceqn}
where
$f_\text{DC}$, $f_\text{SD}$, and $f_\text{C}$ are the relative frequencies of the dark companion, semidetached, and contact binary classes in the sample, respectively, and $R^\text{T}_\text{DC}$, $R^\text{T}_\text{SD}$, and $R^\text{T}_\text{C}$ are the class-specific test recalls of the classifier trained on data with the same noise characteristics as the sample. The expected precision of the classifier for the semidetached and contact binary classes can be calculated analogously. The expected precision $P_\text{DC}$ tells us what fraction of the objects in the sample classified as dark companions we can expect to be actual dark companions, as opposed to $R^\text{T}_\text{DC}$, which tells us what fraction of the actual dark companions in the sample we can expect to be classified as such. The expected precision allows us to estimate the purity of the refined sample of objects classified as dark companions, which is crucial for assessing the cost efficiency of \mbox{follow-up} observations. Alternatively, if the relative frequencies are unknown, we can estimate the prior purity of the sample that is required to achieve a predefined level of purity in the refined sample, helping us decide whether the sample is worth pursuing.

We illustrate the calculation of $P_\text{DC}$ with a model example: consider a sample of ellipsoidal variables with $f_\text{DC}$~$=$~$0.01$, $f_\text{SD}$~$=$~$0.66$, and $f_\text{C}$~$=$~$0.33$, and a classifier with $R^\text{T}_\text{DC}$~$=$~$0.97$, $R^\text{T}_\text{SD}$~$=$~$0.97$, and $R^\text{T}_\text{C}$~$=$~$0.97$, corresponding to the best-case scenario in Table~\ref{tab:results}. If we plug these values into Eq.~\ref{eq:expected_precision}, we obtain $P_\text{DC}$~$\approx$~$0.25$, yielding a refined sample of objects classified as dark companions with an expected purity of approximately $25\%$, which is a significant improvement over the $1\%$ prior purity of the full sample. If we decrease the class recalls to $R^\text{T}_\text{DC}$~$=$~$0.77$, $R^\text{T}_\text{SD}$~$=$~$0.59$, and $R^\text{T}_\text{C}$~$=$~$0.65$, corresponding to the worst-case scenario in Table~\ref{tab:results}, we obtain a refined sample with $P_\text{DC}$~$\approx$~$0.02$, merely doubling the purity of the full sample. This example demonstrates that even low-purity samples can yield significantly improved results if the classifier produces high enough recalls for all classes.

In practice, we cannot reasonably expect that the noise characteristics of real data will exactly match the characteristics of one of our synthetic samples. When designing the synthetic samples, our goal was not to simulate realistic observing conditions but rather to systematically study the effects of correlated and uncorrelated noise on our ability to distinguish between the three binary classes. Consequently, the classifiers trained on the synthetic samples are not directly applicable to real data. However, if we somehow manage to transfer the noise characteristics of real observations to the synthetic data, including the effects of spots and other phenomena that introduce variations from the synthetic models, we can train a classifier on the augmented data and obtain class recalls that are specific to the target sample. In general, this is a challenging task, requiring a detailed understanding of the instrumental noise of the survey and the physical processes affecting the shapes of the light curves, such as flares and pulsations. As a first approximation, we can fit the light curves in the target sample using Gaussian process regression, modeling the mean as a Fourier series of sufficiently high order. We can then inject the residuals from the Fourier fit of a randomly selected light curve into a preselected noiseless synthetic light curve (before normalization), independently repeating the process several times for all light curves in the synthetic data. By training, validating, and testing the classifier on the augmented data, we can obtain class recalls that roughly reflect the noise characteristics of the target sample, allowing us to estimate the expected precision of the classifier on real data.

\section{Discussion and conclusions} \label{sec:conclusions}
In this work, we addressed the issue of whether it is possible to identify noninteracting BHs and NSs in close binary systems based solely on the effects they induce in the broadband photometric light curves of their companion stars. A massive compact companion in a close binary system can tidally deform the primary star into a teardrop shape, causing periodic changes in the area of the star that is visible to the observer and giving rise to ellipsoidal variations in its light curve. By searching for stars that exhibit ellipsoidal variations, we can potentially identify binary systems that host electromagnetically silent BHs and NSs, which we collectively refer to as dark companions. The problem with this approach is that other types of objects, such as contact binaries and semidetached binaries, can also exhibit ellipsoidal variations, making the identification of dark companions challenging.

One way to distinguish dark companion binaries from contaminants is to train a machine learning classifier on a \mbox{well-curated} sample of their observed light curves in which each class is properly represented. However, the limited number of known dark companion binaries prevents us from following this approach. Instead, we generated a large number of synthetic light curves of dark companion binaries, semidetached binaries, and contact binaries, covering a wide range of physical and orbital parameters of the systems (Sect.~\ref{sec:physical_models}). To account for the effects of instrumental noise and stellar spots, we injected the light curves with various levels of correlated and uncorrelated Gaussian noise, resulting in $40$ synthetic samples (Table~\ref{tab:synthetic_samples}). We normalized the light curves in the synthetic samples by fitting them with a fourth-order Fourier series, realigning them to have the primary minimum at phase $0$, and vertically shifting and rescaling them so that their Fourier fits have a minimum and maximum of $0$ and $1$, respectively.

To uncover the underlying discriminative patterns in the high-dimensional synthetic data, we reduced the light curves using PCA and discrete Fourier series -- two linear methods that are well-suited for the decomposition of discrete periodic signals. We performed PCA separately on the noiseless normalized light curves of each binary class, yielding three distinct PCA bases, with the expansion coefficients in the bases forming the PCA representations of the light curves. We also constructed a discretized Fourier basis by sampling the Fourier basis functions on the same grid as the synthetic light curves. For the Fourier representation to be directly comparable with the PCA representations, we subtracted the mean dark companion binary light curve from all light curves, including the semidetached and contact binary light curves, prior to the decomposition. In all four bases, we distinguished between unit and rescaled latent representations, as well as extended unit and rescaled latent representations, where the extended representations contain the amplitude of the light curve as the zeroth component.

Our analysis of the synthetic data revealed that the mean light curves of dark companion binaries, semidetached binaries, and contact binaries are very similar (Fig.~\ref{fig:pca_means}), demonstrating the difficulty of distinguishing between the classes using photometric data alone. The finer details of the light curves are captured by the principal components, which differ between the classes and become progressively more oscillatory with increasing order (Fig.~\ref{fig:pca_components}). In all cases, the cumulative explained variance of the PCA components exceeds $99\%$ somewhere between the third and fifth component (Fig.~\ref{fig:pca_explained_variances}), indicating that the light curves are effectively confined to low-dimensional hyperplanes in the original high-dimensional space and justifying the use of PCA for dimensionality reduction.

Visual inspection of the latent representations of the synthetic samples W0C0, W100C0, W0C10L50, and W100C10L50, which are the corner cases with the lowest and the highest levels of correlated and uncorrelated noise, revealed major differences between the PCA and Fourier representations (Fig.~\ref{fig:scatter_plot_latent_coordinates_1_3_noisy_data_corner_cases}). We found that the first three coefficients of the PCA representations exhibit much richer structure in the latent space than the coefficients of the Fourier representation, in which the distributions of the classes are collapsed along the third coefficient, mixing the classes together and decreasing their separation. This is true for all corner cases, but the class differences are more pronounced in the absence of correlated noise. We found that correlated noise affects the structure of the latent space on a more fundamental level than uncorrelated noise, making it more difficult to visually separate between the classes. Still, even in the presence of strong correlated and uncorrelated noise, the PCA representations retain some of their original structure, while the Fourier representation becomes almost featureless, demonstrating the superiority of the PCA representations.

To compare the class separation in the unit PCA and Fourier representations under different noise conditions, we calculated the silhouette scores of the four corner cases as a function of the number of coefficients in the representation $n$~$=$~$1\text{--}9$ (Sect.~\ref{sec:silhouette_score_results}). Our findings from the analysis of the silhouette scores are largely consistent with the conclusions drawn from the visual inspection of the corner cases. We found that the PCA representations generally yield better class separation than the Fourier representation for a given $n$, with the exception of the first few coefficients, which seem to be more informative in the Fourier representation (Fig.~\ref{fig:silhouette_scores_noisy_data_corner_cases}). This holds true across all corner cases, demonstrating the robustness of the PCA representations to noise. The silhouette scores of the PCA representations typically peak or plateau at $n$~$=$~$3\text{--}6$, possibly with the exception of the contact binary representation, which in some cases exhibits a more gradual increase of the silhouette score with $n$. Other than that, the three PCA representations perform comparably, yielding similar maximum silhouette scores. By comparing the silhouette scores of the latent representations with the benchmark silhouette score of the full representation, we found that the latent representations are much more immune to uncorrelated noise than correlated noise, confirming what we observed in Fig.~\ref{fig:scatter_plot_latent_coordinates_1_3_noisy_data_corner_cases}.

To assess the mean non-overlap of the classes in the latent space, we trained random forest classifiers on the extended rescaled latent representations of the synthetic samples and analyzed their macro recalls as a function of $n$~$=$~$1\text{--}9$ (Sect.~\ref{sec:random_forests_macro_recall_results}). We observed that the macro recalls of the PCA representations start to saturate at $n$~$=$~$3\text{--}6$, depending on the representation and the noise level, and then slightly increase or decrease with increasing $n$ (Fig.~\ref{fig:macro_recalls_corner_cases_only_coeffs}). The saturation points of the PCA representations are slightly shifted to higher $n$ compared to the plateaus and peaks of the silhouette scores, especially in the case of the noiseless sample W0C0. We observed no significant shift in the saturation points of the Fourier representation, indicating good alignment of the silhouette score with the Fourier representation.

Consistent with our analysis of the silhouette scores, we observed that random forest classifiers generally yield higher macro recalls when trained on the PCA representations than the Fourier representation for the same $n$. The Fourier representation does not yield the best macro recall for any of the $40$ synthetic samples, suggesting that the classes are generally more intermixed in the Fourier latent space and pointing to the superiority of the PCA representations in terms of class separation and robustness to noise. (Fig.~\ref{fig:representations_and_dimensions_at_best_recalls}). In most cases, the best macro recalls are obtained for the semidetached binary representation with $n$~$=$~$7\text{--}9$. However, the differences between the three PCA representations are only marginal and the specific choice of the representation does not have a significant effect on the performance of the classifier (Fig.~\ref{fig:macro_recalls_corner_cases_only_coeffs}).

We obtained the best validation macro recalls of the random forest classifiers on the synthetic samples by taking the maximum across all representations and hyperparameter setups (top panel of Fig.~\ref{fig:validation_test_macro_recalls}). By evaluating the best performing random forests on the test sets of the synthetic samples, we verified that the validation macro recalls generalize well to previously unseen data, with a typical decrease of $1\text{--}6\%$ in absolute terms, depending on the synthetic sample (bottom panel of Fig.~\ref{fig:validation_test_macro_recalls}). The test macro recalls of the best performing random forest classifiers vary from $R^\text{T}_\text{M}$~$=$~$0.97$ in the absence of noise to $R^\text{T}_\text{M}$~$=$~$0.67$ in the presence of strong correlated and uncorrelated noise, manifesting low to medium overlap of the classes in the latent space. We found that in the presence of moderate levels of uncorrelated noise ($10^{-4}\text{\,mag}$~$\le$~$\sigma_\text{WN}$~$\le$~$10^{-3}$\,mag), the overlap of the classes is largely determined by the level of correlated noise, with shorter correlation lengths generally yielding worse class separations. Uncorrelated noise starts to significantly affect the macro recalls only at higher levels ($\sigma_\text{WN}$~$>$~$10^{-3}$\,mag), whereas correlated noise can considerably increase the class overlap even at low to moderate levels ($0.01$~$\le$~$\sigma_\text{CN}$~$\le$~$0.05$). This contrast reveals a more fundamental effect of correlated noise on the separation of the classes in the latent space. Nevertheless, even in the presence of strong correlated noise ($\sigma_\text{CN}$~$=$~$0.1$), which can amount to significant surface coverage with stellar spots, the classes remain largely separated in the latent space, with test macro recalls reaching $R^\text{T}_\text{M}$~$=$~$0.67\text{--}0.71$.

We retrained the random forest classifiers on the extended representations augmented with the variances of the photometric amplitude and the latent coefficients to investigate whether the inclusion of variances in the input of the classifiers improves the separation of the classes (Sect.~\ref{sec:impact_of_variances}). We found that while the macro recalls of \mbox{low-dimensional} PCA representations ($n$~$\lesssim$~$3$) significantly increase, the improvement is only marginal beyond the saturation points of the unaugmented representations at $n$~$=$~$3\text{--}6$. The positive effect of the variances is the most pronounced in the case of the Fourier representation, where the macro recalls are considerably improved for $n$~$=$~$1\text{--}9$ across all noise conditions (Fig.~\ref{fig:macro_recalls_corner_cases_augmented_representations}), even surpassing the macro recalls of the PCA representations in some cases (Fig.~\ref{fig:augmented_representations_and_dimensions_at_best_recalls}). However, the best macro recalls of the random forest classifiers trained on the augmented representations (Fig.~\ref{fig:validation_test_macro_recalls_augmented_representations}) remain largely unchanged compared to the macro recalls obtained for the unaugmented representations, pointing to the limited benefit of including variances in the representations.

Using the obtained test class recalls (Table~\ref{tab:results}), we showed that it is possible to estimate the expected precision of the classifier on real data, assuming we have a rough estimate of the relative frequencies of the classes in the sample (Sect.~\ref{sec:expected_precision}). We illustrated the calculation of the expected precision on a model example, and we showed that in the best-case scenario, we can increase the purity of a dark companion sample by a factor of up to $25$, assuming a prior purity of $1\%$.

There are several limitations to our study that need to be addressed. First, we generated the synthetic light curves using PHOEBE binary models, which are not perfect representations of reality. As a result, the synthetic light curves may not capture all the complexities of real light curves, such as Doppler beaming and boosting \citep{loeb03,zucker07}, which are not supported as of version $2.2$. In addition, we made several simplifying assumptions in the generation of the synthetic light curves, such as the circularity of the orbits or the default limb darkening calculation settings. However, we expect that these effects are secondary to the main features of the light curves and become negligible in the presence of noise. Consequently, we do not expect the main findings of our study, such as the superiority of the PCA representations over the Fourier representation, to be significantly affected by these assumptions.

Second, we injected the synthetic light curves with various levels of uncorrelated and correlated Gaussian noise to account for the effects of instrumental noise and stellar spots, respectively. While the uncorrelated noise is a good approximation of the instrumental noise, the correlated noise is a very simplified model of the effects of stellar spots, which can be more complex in reality. Also, we limited our analysis to $l_\text{CN}$~$=$~$0.25$, $0.5$, and $1$, which may not cover the full range of possible timescales of the spots. A more realistic treatment of spots could be achieved either by simulating the spots directly in PHOEBE, which would greatly increase the size and complexity of the synthetic data as well as the computational cost of its generation, or by using spot models of single stars \citep[e.g.,][]{luger19,luger21a,luger21b} instead of correlated Gaussian noise. In addition, we did not consider the effects of other sources of noise, such as the intrinsic variability of the stars. All these effects can potentially complicate the separation of the classes in the latent space and decrease the macro recalls on real data. To avoid modeling the noise altogether, we can transfer the noise characterics directly from the target sample as we described in Sect.~\ref{sec:expected_precision}, taking into account all the complexities and idiosyncracies of real astronomical observations and yielding a more accurate estimate of the expected precision of the classifier on real data. There are several samples of ellipsoidal variables in the literature that are suitable for this purpose \citep[e.g.,][]{Green_2023,Gomel_2023,Gomel_2021_ogle}. We plan to investigate these samples using our method in future work.

Third, in our analysis, we implicitly assumed that ellipsoidal samples contain only dark companion binaries, semidetached binaries, and contact binaries, which we regard as the most challenging classes to separate. In reality, low-inclination detached binaries are also prominently present in ellipsoidal samples, and other types of objects, such as pulsating stars and spotted rotating stars, can be found in the samples as well. We excluded detached binaries from our analysis, because we assumed that their light curves are sufficiently similar to those of semidetached binaries to be treated as such by the classifier, effectively increasing the relative frequency of the semidetached class in the sample. Although further investigation is required to confirm the validity of this assumption, our analysis of the latent representations of the TESS ellipsoidal sample from \citet{Green_2023} revealed a significant overlap of this sample with our synthetic data (Fig.~\ref{fig:scatter_plot_coefficients_1_3_TESS_ellipsoidal_variables}), suggesting that the assumption is reasonable. Still, our method can be easily extended to accommodate detached binaries as a separate class if needed, allowing for a more detailed differentiation between the classes. As for pulsating stars and spotted rotating stars, these can be efficiently removed from the sample by performing suitable cuts on period and amplitude \citep{Green_2023}. Consequently, we assume that the fraction of these objects in ellipsoidal samples is negligible, and their effect on the performance of the classifiers is minimal. Lastly, in our definition of dark companion binaries, the dark companion is either a BH or a NS. In practice, it is difficult to distinguish ellipsoidal variations induced by a NS from those induced by a massive white dwarf (WD), leading to contamination of the dark companion class by these objects. Since the true nature of the dark companion can only be reliably determined through high-resolution spectroscopy, we did not attempt to separate dark companion binaries from binaries hosting massive WDs. Instead, we treated them as a single class in our analysis, leaving their separation up to \mbox{follow-up} observations. To assess the level of contamination by massive WDs, further analysis of the candidates identified by our method is required.

Fourth, we trained the classifiers on the extended latent representations, which encode the absolute scales and the morphologies of the light curves, but do not take into account their periods. While we expect that most of the discriminative information is contained in the shapes of the light curves, the periods can provide additional information about the physical characters of the systems, potentially improving the separation of the classes. We plan to investigate the impact of including the periods in the latent representations on the performance of the classifiers in future work.

Fifth, to obtain the boundaries and quantify the overlap between the classes, we trained the random forest classifiers on the oversampled synthetic data, which we further balanced by weighting the objects with the inverse of the class size. This is a valid approach, provided the distributions of the objects within the classes are representative. That is, the objects of a given class populate the parameter space in a way that is representative of real data. However, this is not the case in our synthetic data, which we generated on uniform and \mbox{log-uniform} grids of physical and orbital parameters of the systems (Sect.~\ref{sec:physical_models}). Our motivation was to cover a wide range of parameters in order to capture the full diversity of the light curves rather than to mimic the real distributions of the objects. Consequently, the boundaries of the classes in the latent space may be distorted with respect to real data, leading to biased estimates of the macro recalls to either side. This can only be avoided by generating synthetic light curves with representative distributions of the parameters, which is a challenging task given the complexity of the parameter space. Representative distributions of at least some physical parameters could be obtained using binary population synthesis \citep[e.g.,][]{Weller_2023,Chawla_2024}. Until we train the classifiers on data with representative parameter distributions, we cannot quantify the impact of parameter sampling strategy on the macro recalls and the expected precision of the classifiers. However, assuming that the objects within the classes do not accumulate close to the decision boundaries in the latent space, we expect our estimates to generalize well to real data.

Our method of projecting light curves to PCA components learned from synthetic data can be easily extended to multiband photometry, which can provide additional information about the physical properties of the systems. For example, ultraviolet photometry from upcoming satellites such as QUVIK \citep{Werner_2024,Krticka_2024} or ULTRASAT \citep{Shvartzvald_2024} could be used to constrain the nature of semidetached binaries, including binaries with stripped-envelope stars \citep{Rowan_2024}. Precise ultraviolet observations could also be used to further break the degeneracy between contact binaries and dark companion binaries using the different limb darkening properties of these systems in UV. We plan to pursue this direction in future work.

Another possibility of improving our method is to explicitly model the correlated noise in the light curves using Gaussian processes. We found that correlated noise affects the light curves on a more fundamental level than uncorrelated noise, preventing the latent representations from disentangling the signal from the noise. This is not suprising -- by fitting the light curves using least squares, we implicitly assume homoscedastic uncorrelated Gaussian noise, which is clearly not justified in the presence of correlated noise. We can avoid this assumption by modeling the light curves using Gaussian process regression with a nonzero mean given by a linear combination of Fourier components. We expect that this approach will yield more robust latent representations of the light curves and better separation of the classes in the latent space. The recovered parameters of the correlated noise can possibly also be informative about the physical properties of the systems. We intend to explore this approach in future work.

\begin{acknowledgements}
MP thanks Yuan-Sen Ting for helpful discussions. We thank the anonymous referee for their constructive comments that helped improve the quality of the paper. We acknowledge the support of the Czech Science Foundation Grant No. 24-11023S. This work made use of the following software packages: \texttt{PHOEBE} \citep{Prsa_2016,Conroy_2020}, \texttt{numpy} \citep{numpy}, \texttt{scikit-learn} \citep{scikit_learn}, \texttt{matplotlib} \citep{matplotlib}, \texttt{pandas} \citep{pandas}.
\end{acknowledgements}

\bibliographystyle{aa}
\bibliography{bibliography}

\onecolumn
\begin{appendix}
\section{Synthetic samples} \label{app:synthetic_samples}
\begin{table}[h!]
    \caption{Synthetic samples of dark companion, semidetached, and contact binary light curves.}
    \label{tab:synthetic_samples}
    \begin{center}
    \begin{tabular}{lccc}
    \hline\hline
    Sample & $\sigma_\text{WN}$ (mag) & $\sigma_\text{CN}$ & $l_\text{CN}$\\
    \hline
    W0C0 & -- & -- & --\\
    W0C1L25 & -- & $0.01$ & $0.25$\\
    W0C1L50 & -- & $0.01$ & $0.50$\\
    W0C1L100 & -- & $0.01$ & $1.00$\\
    W0C5L25 & -- & $0.05$ & $0.25$\\
    W0C5L50 & -- & $0.05$ & $0.50$\\
    W0C5L100 & -- & $0.05$ & $1.00$\\
    W0C10L25 & -- & $0.10$ & $0.25$\\
    W0C10L50 & -- & $0.10$ & $0.50$\\
    W0C10L100 & -- & $0.10$ & $1.00$\\
    W1C0 & $10^{-4}$ & -- & --\\
    W1C1L25 & $10^{-4}$ & $0.01$ & $0.25$\\
    W1C1L50 & $10^{-4}$ & $0.01$ & $0.50$\\
    W1C1L100 & $10^{-4}$ & $0.01$ & $1.00$\\
    W1C5L25 & $10^{-4}$ & $0.05$ & $0.25$\\
    W1C5L50 & $10^{-4}$ & $0.05$ & $0.50$\\
    W1C5L100 & $10^{-4}$ & $0.05$ & $1.00$\\
    W1C10L25 & $10^{-4}$ & $0.10$ & $0.25$\\
    W1C10L50 & $10^{-4}$ & $0.10$ & $0.50$\\
    W1C10L100 & $10^{-4}$ & $0.10$ & $1.00$\\
    W10C0 & $10^{-3}$ & -- & --\\
    W10C1L25 & $10^{-3}$ & $0.01$ & $0.25$\\
    W10C1L50 & $10^{-3}$ & $0.01$ & $0.50$\\
    W10C1L100 & $10^{-3}$ & $0.01$ & $1.00$\\
    W10C5L25 & $10^{-3}$ & $0.05$ & $0.25$\\
    W10C5L50 & $10^{-3}$ & $0.05$ & $0.50$\\
    W10C5L100 & $10^{-3}$ & $0.05$ & $1.00$\\
    W10C10L25 & $10^{-3}$ & $0.10$ & $0.25$\\
    W10C10L50 & $10^{-3}$ & $0.10$ & $0.50$\\
    W10C10L100 & $10^{-3}$ & $0.10$ & $1.00$\\
    W100C0 & $10^{-2}$ & -- & --\\
    W100C1L25 & $10^{-2}$ & $0.01$ & $0.25$\\
    W100C1L50 & $10^{-2}$ & $0.01$ & $0.50$\\
    W100C1L100 & $10^{-2}$ & $0.01$ & $1.00$\\
    W100C5L25 & $10^{-2}$ & $0.05$ & $0.25$\\
    W100C5L50 & $10^{-2}$ & $0.05$ & $0.50$\\
    W100C5L100 & $10^{-2}$ & $0.05$ & $1.00$\\
    W100C10L25 & $10^{-2}$ & $0.10$ & $0.25$\\
    W100C10L50 & $10^{-2}$ & $0.10$ & $0.50$\\
    W100C10L100 & $10^{-2}$ & $0.10$ & $1.00$\\
    \hline
    \end{tabular}
    \tablefoot{Each sample contains a total of $42\,700$ light curves ($11\,220$ dark companion, $18\,460$ semidetached, and $13\,020$ contact binaries) generated with a different combination of uncorrelated noise standard deviation $\sigma_\text{WN}$, correlated noise standard deviation $\sigma_\text{CN}$ (proportional to the unperturbed light curve amplitude), and correlation length scale $l_\text{CN}$ (in units of the orbital period).}
    \end{center}
\end{table}

\section{Silhouette score} \label{app:silhouette_score}
The silhouette score, which quantifies how similar an object is to its own class compared to the other classes, is one of the most popular clustering measures \citep{Rousseeuw_1987}. The score ranges from $-1$ to $1$, with higher values indicating better separation of the classes. The silhouette score of the $i$th object in the class $K$ is calculated as
\begin{ceqn}
\begin{equation}
    s^{K}_{i} = \frac{b^{K^\prime}_i - a^{K}_{i}}{\max(a^{K}_{i}, b^{K^\prime}_{i})},
\end{equation}
\end{ceqn}
where $a^{K}_{i}$ is the average Euclidean distance of the $i$th object to all other objects in the class $K$, and $b^{K^\prime}_{i}$ is the average distance of the $i$th object to all objects in the closest neighboring class $K^\prime$~$\ne$~$K$. In our analysis, we utilized the \texttt{scikit-learn} implementation of the silhouette score, namely the \texttt{silhouette\_samples} function, which returns the silhouette score of each object in the dataset. We calculated the overall silhouette score of the dataset as the average of the individual silhouette scores weighted by the inverse of the respective class sizes, yielding a robust measure of clustering quality that is insensitive to class imbalance.

In general, higher values of the silhouette score indicate better separation of the classes in the dataset. However, the silhouette score is not invariant under independent rescaling of the features, meaning that the score can be artificially inflated by rescaling each feature with a different factor. Without a physically motivated scaling of the features, the absolute value of the silhouette score is not meaningful. Still, the silhouette scores of two competing representations can be directly compared to asses which representation separates the classes better, provided the bases of the representations are brought to the same (arbitrary) scale. The simplest way to achieve this is to normalize the basis vectors to have unit norms, which is the approach we followed in this work.

\section{Macro recall} \label{app:macro_recall}
In a classification task, the recall $R_{K}$ of a classifier for the class $K$ is defined as
\begin{ceqn}
\begin{equation}
    R_{K} = \frac{TP_{K}}{TP_{K} + FN_{K}},
    \label{eq:recall}
\end{equation}
\end{ceqn}
where $TP_{K}$ is the number of class-$K$ true positives (correctly predicted objects in the class $K$) and $FN_{K}$ is the number of class-$K$ false negatives (objects in the class $K$ that were incorrectly predicted as belonging to a different class). The macro recall $R_\text{M}$ of the classifier is calculated as a simple arithmetic average of the recalls of the $N_\text{classes}$ individual classes,
\begin{ceqn}
\begin{equation}
    R_\text{M} = \frac{1}{N_\text{classes}}\sum_{K=1}^{N_\text{classes}} R_{K}.
    \label{eq:macro_recall}
\end{equation}
\end{ceqn}
The benefit of the macro recall is that it is not sensitive to the class sizes, which makes it an ideal measure of class separation for imbalanced datasets or datasets with unknown class frequencies, such as the synthetic data in this work. The macro recall should not be confused with the accuracy $A$ of the classifier, which is calculated as
\begin{ceqn}
\begin{equation}
    A = \frac{\sum_{K=1}^{N_\text{classes}} TP_{K}}{\sum_{K=1}^{N_\text{classes}} (TP_{K} + FN_{K})} = \sum_{K=1}^{N_\text{classes}} f_{K} R_{K},
    \label{eq:accuracy}
\end{equation}
\end{ceqn}
where $f_{K}$ are the relative frequencies of the classes in the dataset. By comparing Eqs.~(\ref{eq:accuracy}) and (\ref{eq:macro_recall}), we see that the macro recall coincides with the accuracy only if the classes are balanced, that is, if $f_{K}$~$=$~$1/N_\text{classes}$ for all $K$. In the case of imbalanced classes, the accuracy is skewed toward the recall of the majority class, while the macro recall treats all classes with equal importance. However, if we train the classifier directly on the imbalanced data, the macro recall can also become skewed, provided the classes are not well-separated in the latent space. For the macro recall to be a robust measure of the mean non-overlap of the classes in the latent space, we need to artificially balance the data by weighing the samples with the inverse of their class sizes prior to training the classifier. This way, we can ensure that the classifier is not biased toward the majority class and the class contours in the latent space are not affected by the class sizes.
\clearpage
\newpage

\section{Scatter plots of latent representations of synthetic samples} \label{app:scatter_plots}
\vspace*{\fill}
\begin{figure*}[h!]
    \centering
    \includegraphics[width=0.95\textwidth]{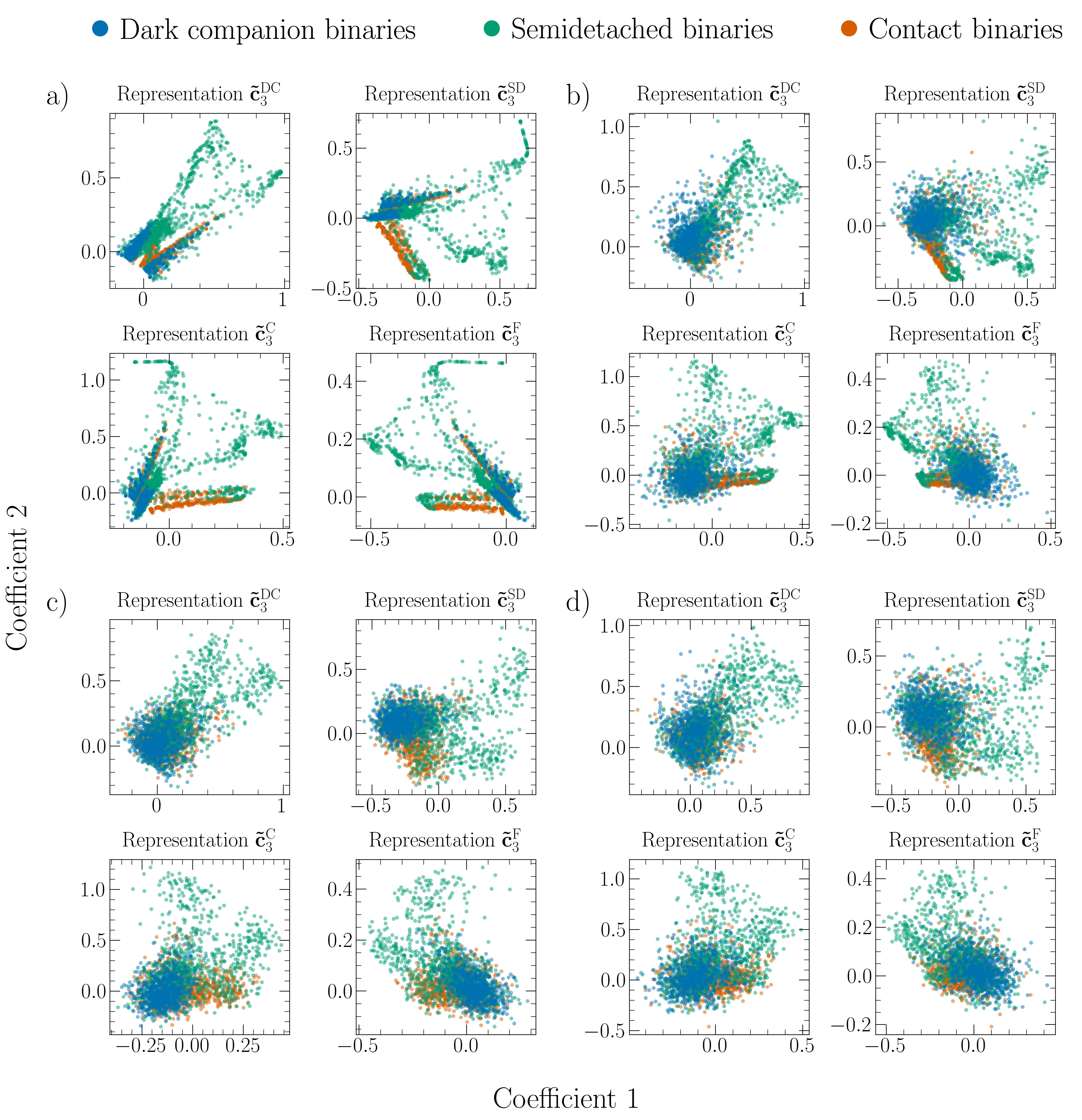}
    \caption{Scatter plots of the first and second coefficients of the representations $\mathbf{\tilde{c}^\text{DC}}$, $\mathbf{\tilde{c}^\text{SD}}$, $\mathbf{\tilde{c}^\text{C}}$, and $\mathbf{\tilde{c}^\text{F}}$ of the dark companion, semidetached, and contact binary light curves in the validation sets of the synthetic samples W0C0 (a), W100C0 (b), W0C10L50 (c), and W100C10L50 (d). We describe the synthetic samples in Sect.~\ref{sec:data} and provide the definitions of the representations in Sects.~\ref{sec:pca_representations}--\ref{sec:fourier_representation}.
    \label{fig:scatter_plot_latent_coordinates_1_2_noisy_data_corner_cases}}
\end{figure*}
\vspace{\fill}
\clearpage
\newpage

\vspace*{\fill}
\begin{figure*}[h!]
    \centering
    \includegraphics[width=0.95\textwidth]{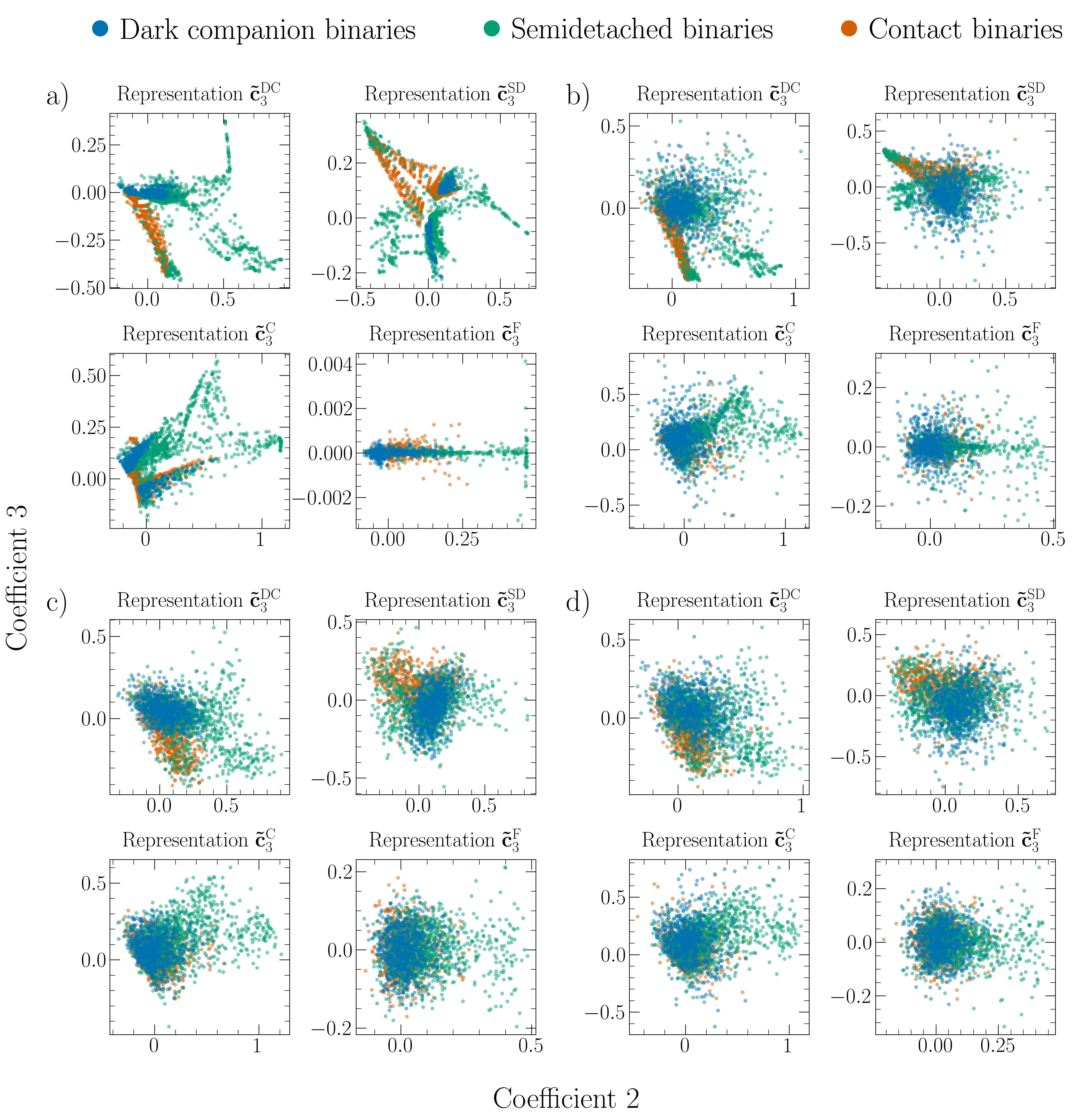}
    \caption{Scatter plots of the second and third coefficients of the representations $\mathbf{\tilde{c}^\text{DC}}$, $\mathbf{\tilde{c}^\text{SD}}$, $\mathbf{\tilde{c}^\text{C}}$, and $\mathbf{\tilde{c}^\text{F}}$ of the dark companion, semidetached, and contact binary light curves in the validation sets of the synthetic samples W0C0 (a), W100C0 (b), W0C10L50 (c), and W100C10L50 (d). We describe the synthetic samples in Sect.~\ref{sec:data} and provide the definitions of the representations in Sects.~\ref{sec:pca_representations}--\ref{sec:fourier_representation}.
    \label{fig:scatter_plot_latent_coordinates_2_3_noisy_data_corner_cases}}
\end{figure*}
\vspace{\fill}
\clearpage
\newpage

\section{Scatter plots of latent representations of TESS ellipsoidal sample}
\vspace*{\fill}

\begin{figure*}[h!]
    \captionsetup[subfigure]{labelformat=empty}
    \centering
    \begin{subfigure}{\textwidth}
        \centering
        \includegraphics[width=\textwidth]{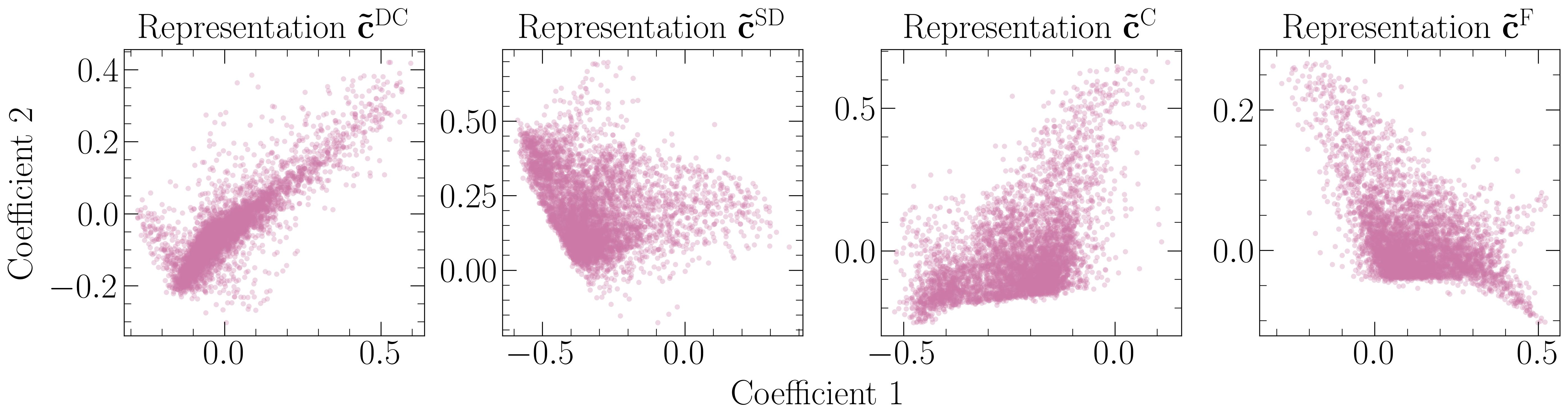}
        \caption{\label{fig:scatter_plot_latent_coordinates_1_2_TESS_ellipsoidal_variables}}
    \end{subfigure}
    
    \vspace{0.5cm}
    
    \begin{subfigure}{\textwidth}
        \centering
        \includegraphics[width=\textwidth]{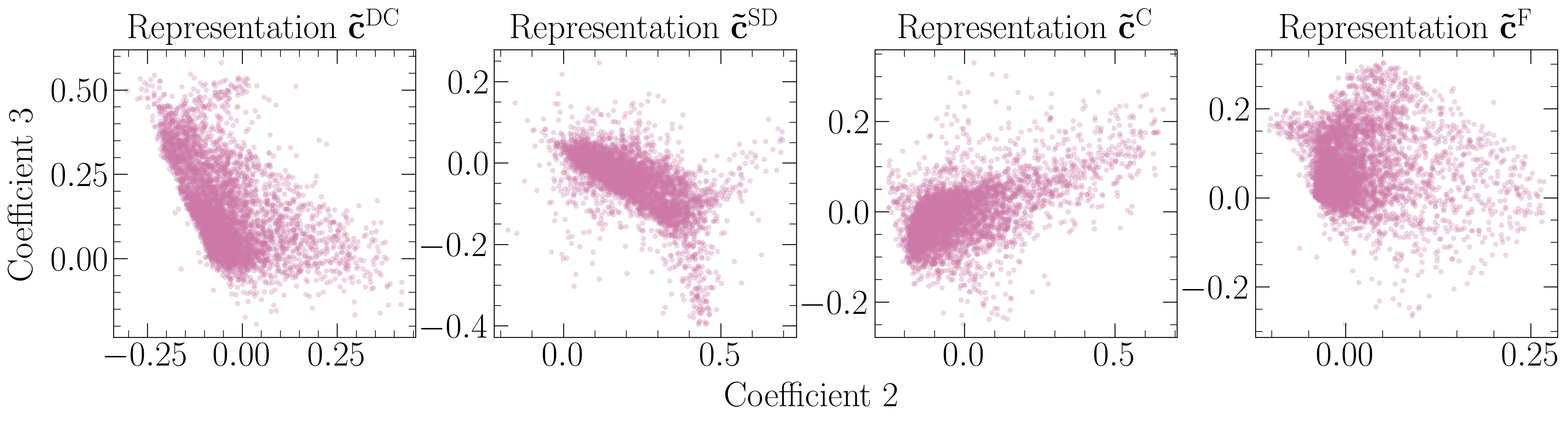}
        \caption{\label{fig:scatter_plot_latent_coordinates_2_3_TESS_ellipsoidal_variables}}
    \end{subfigure}
    
    \caption{Scatter plots of the first and second coefficients (top row) and the second and third coefficients (bottom row) of the representations $\mathbf{\tilde{c}^\text{DC}}$, $\mathbf{\tilde{c}^\text{SD}}$, $\mathbf{\tilde{c}^\text{C}}$, and $\mathbf{\tilde{c}^\text{F}}$ of the TESS ellipsoidal sample from \citet{Green_2023}. We provide the definitions of the representations in Sects.~\ref{sec:pca_representations}--\ref{sec:fourier_representation}.}
    \label{fig:scatter_plot_latent_coordinates_TESS_ellipsoidal_variables}
\end{figure*}
\vspace{\fill}
\clearpage
\newpage

\section{Optimal representations and random forest hyperparameters} \label{app:optimal_representations_hyperparameters}
\begin{table*}[ht!]
    \caption{
    Optimal representations and random forest hyperparameters that yielded the best validation macro recalls ($R^\text{V}_\text{M}$) on the synthetic samples.}
    \label{tab:results}
    \begin{center}
    \resizebox{\textwidth}{!}{
    \renewcommand{\arraystretch}{1.4}
    \begin{tabular}{lcccccccccccc}
    \hline\hline
    Sample & Representation & \texttt{n\_estimators} & \texttt{min\_samples\_leaf} & \texttt{max\_features} & $R^\text{V}_\text{DC}$ & $R^\text{V}_\text{SD}$ & $R^\text{V}_\text{C}$ & $R^\text{V}_\text{M}$ & $R^\text{T}_\text{DC}$ & $R^\text{T}_\text{SD}$ & $R^\text{T}_\text{C}$ & $R^\text{T}_\text{M}$\\[0.5ex]
    \hline
    W0C0        & $\mathbf{\tilde{C}}^\text{DC}_{8}$             & $100$ & $1$  & sqrt & $0.98$ & $0.97$ & $0.98$ & $0.98$ & $0.97$ & $0.97$ & $0.97$ & $0.97$\\
    W0C1L25     & $\mathbf{\tilde{C}}^\text{SD}_{7}$             & $500$ & $1$  & sqrt & $0.90$ & $0.89$ & $0.96$ & $0.92$ & $0.87$ & $0.87$ & $0.95$ & $0.89$\\
    W0C1L50     & $\mathbf{\tilde{C}}^\text{SD}_{9}$             & $100$ & $1$  & sqrt & $0.93$ & $0.92$ & $0.96$ & $0.94$ & $0.91$ & $0.90$ & $0.96$ & $0.92$\\
    W0C1L100    & $\mathbf{\tilde{C}}^\text{SD}_{8}$             & $500$ & $1$  & sqrt & $0.96$ & $0.96$ & $0.97$ & $0.96$ & $0.95$ & $0.96$ & $0.97$ & $0.96$\\
    W0C5L25     & $\mathbf{\tilde{C}}^\text{C\hphantom{D}}_{9}$  & $100$ & $10$ & sqrt & $0.85$ & $0.74$ & $0.85$ & $0.81$ & $0.81$ & $0.68$ & $0.81$ & $0.77$\\
    W0C5L50     & $\mathbf{\tilde{C}}^\text{SD}_{9}$             & $500$ & $10$ & sqrt & $0.84$ & $0.76$ & $0.86$ & $0.82$ & $0.82$ & $0.71$ & $0.82$ & $0.78$\\
    W0C5L100    & $\mathbf{\tilde{C}}^\text{SD}_{9}$             & $500$ & $1$  & sqrt & $0.89$ & $0.90$ & $0.94$ & $0.91$ & $0.87$ & $0.87$ & $0.91$ & $0.89$\\
    W0C10L25    & $\mathbf{\tilde{C}}^\text{C\hphantom{D}}_{9}$  & $500$ & $10$ & sqrt & $0.80$ & $0.67$ & $0.74$ & $0.74$ & $0.76$ & $0.61$ & $0.70$ & $0.69$\\
    W0C10L50    & $\mathbf{\tilde{C}}^\text{SD}_{9}$             & $500$ & $10$ & sqrt & $0.76$ & $0.71$ & $0.72$ & $0.73$ & $0.75$ & $0.66$ & $0.70$ & $0.70$\\
    W0C10L100   & $\mathbf{\tilde{C}}^\text{LC}_{99}$            & $100$ & $10$ & sqrt & $0.88$ & $0.76$ & $0.91$ & $0.85$ & $0.84$ & $0.73$ & $0.89$ & $0.82$\\
    W1C0        & $\mathbf{\tilde{C}}^\text{DC}_{9}$             & $500$ & $1$  & sqrt & $0.96$ & $0.95$ & $0.98$ & $0.96$ & $0.96$ & $0.94$ & $0.97$ & $0.96$\\
    W1C1L25     & $\mathbf{\tilde{C}}^\text{SD}_{8}$             & $500$ & $10$ & sqrt & $0.93$ & $0.86$ & $0.96$ & $0.92$ & $0.90$ & $0.84$ & $0.95$ & $0.89$\\
    W1C1L50     & $\mathbf{\tilde{C}}^\text{SD}_{9}$             & $500$ & $10$ & sqrt & $0.94$ & $0.89$ & $0.97$ & $0.93$ & $0.93$ & $0.88$ & $0.95$ & $0.92$\\
    W1C1L100    & $\mathbf{\tilde{C}}^\text{SD}_{9}$             & $500$ & $1$  & sqrt & $0.95$ & $0.94$ & $0.98$ & $0.96$ & $0.95$ & $0.93$ & $0.97$ & $0.95$\\
    W1C5L25     & $\mathbf{\tilde{C}}^\text{C\hphantom{D}}_{6}$  & $500$ & $10$ & sqrt & $0.85$ & $0.74$ & $0.85$ & $0.81$ & $0.82$ & $0.67$ & $0.81$ & $0.77$\\
    W1C5L50     & $\mathbf{\tilde{C}}^\text{SD}_{9}$             & $100$ & $10$ & sqrt & $0.84$ & $0.76$ & $0.85$ & $0.82$ & $0.82$ & $0.72$ & $0.80$ & $0.78$\\
    W1C5L100    & $\mathbf{\tilde{C}}^\text{SD}_{9}$             & $500$ & $1$  & sqrt & $0.89$ & $0.88$ & $0.93$ & $0.90$ & $0.87$ & $0.86$ & $0.91$ & $0.88$\\
    W1C10L25    & $\mathbf{\tilde{C}}^\text{SD}_{7}$             & $100$ & $10$ & sqrt & $0.79$ & $0.67$ & $0.73$ & $0.73$ & $0.78$ & $0.61$ & $0.69$ & $0.69$\\
    W1C10L50    & $\mathbf{\tilde{C}}^\text{SD}_{9}$             & $100$ & $10$ & sqrt & $0.76$ & $0.69$ & $0.74$ & $0.73$ & $0.74$ & $0.65$ & $0.69$ & $0.69$\\
    W1C10L100   & $\mathbf{\tilde{C}}^\text{SD}_{9}$             & $500$ & $1$  & sqrt & $0.84$ & $0.84$ & $0.86$ & $0.85$ & $0.79$ & $0.81$ & $0.82$ & $0.81$\\
    W10C0       & $\mathbf{\tilde{C}}^\text{SD}_{8}$             & $500$ & $1$  & sqrt & $0.92$ & $0.89$ & $0.96$ & $0.92$ & $0.90$ & $0.87$ & $0.93$ & $0.90$\\
    W10C1L25    & $\mathbf{\tilde{C}}^\text{SD}_{8}$             & $100$ & $10$ & sqrt & $0.92$ & $0.83$ & $0.95$ & $0.90$ & $0.91$ & $0.80$ & $0.91$ & $0.88$\\
    W10C1L50    & $\mathbf{\tilde{C}}^\text{DC}_{6}$             & $500$ & $1$  & sqrt & $0.90$ & $0.87$ & $0.94$ & $0.90$ & $0.89$ & $0.84$ & $0.91$ & $0.88$\\
    W10C1L100   & $\mathbf{\tilde{C}}^\text{SD}_{9}$             & $500$ & $1$  & sqrt & $0.91$ & $0.88$ & $0.95$ & $0.91$ & $0.89$ & $0.86$ & $0.92$ & $0.89$\\
    W10C5L25    & $\mathbf{\tilde{C}}^\text{C\hphantom{D}}_{9}$  & $100$ & $10$ & sqrt & $0.86$ & $0.74$ & $0.85$ & $0.82$ & $0.83$ & $0.67$ & $0.79$ & $0.76$\\
    W10C5L50    & $\mathbf{\tilde{C}}^\text{SD}_{9}$             & $500$ & $10$ & sqrt & $0.85$ & $0.76$ & $0.84$ & $0.82$ & $0.82$ & $0.70$ & $0.81$ & $0.78$\\
    W10C5L100   & $\mathbf{\tilde{C}}^\text{DC}_{9}$             & $500$ & $1$  & sqrt & $0.85$ & $0.84$ & $0.90$ & $0.86$ & $0.82$ & $0.81$ & $0.87$ & $0.83$\\
    W10C10L25   & $\mathbf{\tilde{C}}^\text{C\hphantom{D}}_{6}$  & $100$ & $10$ & sqrt & $0.76$ & $0.67$ & $0.73$ & $0.72$ & $0.74$ & $0.61$ & $0.69$ & $0.68$\\
    W10C10L50   & $\mathbf{\tilde{C}}^\text{SD}_{9}$             & $500$ & $10$ & sqrt & $0.76$ & $0.70$ & $0.75$ & $0.74$ & $0.74$ & $0.65$ & $0.69$ & $0.69$\\
    W10C10L100  & $\mathbf{\tilde{C}}^\text{SD}_{8}$             & $500$ & $10$ & None  & $0.84$ & $0.75$ & $0.84$ & $0.81$ & $0.82$ & $0.70$ & $0.80$ & $0.77$\\
    W100C0      & $\mathbf{\tilde{C}}^\text{C\hphantom{D}}_{5}$  & $500$ & $10$ & sqrt & $0.90$ & $0.73$ & $0.81$ & $0.82$ & $0.89$ & $0.69$ & $0.77$ & $0.78$\\
    W100C1L25   & $\mathbf{\tilde{C}}^\text{C\hphantom{D}}_{4}$  & $100$ & $10$ & sqrt & $0.89$ & $0.73$ & $0.82$ & $0.81$ & $0.86$ & $0.68$ & $0.78$ & $0.77$\\
    W100C1L50   & $\mathbf{\tilde{C}}^\text{SD}_{7}$             & $500$ & $10$ & log2 & $0.91$ & $0.74$ & $0.81$ & $0.82$ & $0.87$ & $0.68$ & $0.77$ & $0.77$\\
    W100C1L100  & $\mathbf{\tilde{C}}^\text{C\hphantom{D}}_{5}$  & $100$ & $10$ & sqrt & $0.89$ & $0.73$ & $0.81$ & $0.81$ & $0.87$ & $0.68$ & $0.77$ & $0.77$\\
    W100C5L25   & $\mathbf{\tilde{C}}^\text{SD}_{7}$             & $500$ & $10$ & log2 & $0.85$ & $0.70$ & $0.77$ & $0.78$ & $0.82$ & $0.65$ & $0.72$ & $0.73$\\
    W100C5L50   & $\mathbf{\tilde{C}}^\text{SD}_{6}$             & $100$ & $10$ & sqrt & $0.83$ & $0.71$ & $0.77$ & $0.77$ & $0.80$ & $0.66$ & $0.73$ & $0.73$\\
    W100C5L100  & $\mathbf{\tilde{C}}^\text{SD}_{7}$             & $500$ & $10$ & log2 & $0.82$ & $0.72$ & $0.79$ & $0.78$ & $0.79$ & $0.68$ & $0.74$ & $0.74$\\
    W100C10L25  & $\mathbf{\tilde{C}}^\text{DC}_{8}$             & $500$ & $10$ & sqrt & $0.80$ & $0.67$ & $0.68$ & $0.72$ & $0.77$ & $0.59$ & $0.65$ & $0.67$\\
    W100C10L50  & $\mathbf{\tilde{C}}^\text{DC}_{6}$             & $500$ & $10$ & sqrt & $0.78$ & $0.68$ & $0.68$ & $0.71$ & $0.76$ & $0.62$ & $0.65$ & $0.68$\\
    W100C10L100 & $\mathbf{\tilde{C}}^\text{SD}_{8}$             & $500$ & $10$ & sqrt & $0.78$ & $0.71$ & $0.74$ & $0.75$ & $0.76$ & $0.66$ & $0.71$ & $0.71$\\
    \hline
    \end{tabular}
    }
    \tablefoot{We also present the validation class recalls for the dark companion ($R^\text{V}_\text{DC}$), semidetached ($R^\text{V}_\text{SD}$), and contact ($R^\text{V}_\text{C}$) binary light curves as well as the test class recalls ($R^\text{T}_\text{DC}$, $R^\text{T}_\text{SD}$, $R^\text{T}_\text{C}$) and the test macro recall ($R^\text{T}_\text{M}$) of the best performing classifiers. We describe the synthetic samples in Sect.~\ref{sec:data} and provide the definitions of the representations in Sects.~\ref{sec:pca_representations}--\ref{sec:fourier_representation}. The symbol $\mathbf{\tilde{C}}^\text{LC}_{99}$ denotes the extended full representation. The definitions of the hyperparameters are given in Sect.~\ref{sec:random_forests_macro_recall_methods}.}
    \end{center}
\end{table*}

\begin{table*}[ht!]
    \caption{
    Optimal augmented representations and random forest hyperparameters that yielded the best validation macro recalls ($R^\text{V}_\text{M}$) on the synthetic samples.}
    \label{tab:results_augmented_representations}
    \begin{center}
    \resizebox{\textwidth}{!}{
    \renewcommand{\arraystretch}{1.4}
    \begin{tabular}{lcccccccccccc}
    \hline\hline
    Sample & Representation & \texttt{n\_estimators} & \texttt{min\_samples\_leaf} & \texttt{max\_features} & $R^\text{V}_\text{DC}$ & $R^\text{V}_\text{SD}$ & $R^\text{V}_\text{C}$ & $R^\text{V}_\text{M}$ & $R^\text{T}_\text{DC}$ & $R^\text{T}_\text{SD}$ & $R^\text{T}_\text{C}$ & $R^\text{T}_\text{M}$\\[0.5ex]
    \hline
    W0C0        & $\mathbf{\tilde{C}}^\text{F+V\hphantom{D}}_{8}$  & $100$ & $1$  & sqrt & $0.97$ & $0.99$ & $0.98$ & $0.98$ & $0.96$ & $0.97$ & $0.97$ & $0.97$\\
    W0C1L25     & $\mathbf{\tilde{C}}^\text{SD+V}_{8}$             & $500$ & $1$  & sqrt & $0.90$ & $0.89$ & $0.97$ & $0.92$ & $0.87$ & $0.86$ & $0.95$ & $0.89$\\
    W0C1L50     & $\mathbf{\tilde{C}}^\text{SD+V}_{9}$             & $500$ & $1$  & sqrt & $0.93$ & $0.92$ & $0.96$ & $0.93$ & $0.90$ & $0.90$ & $0.95$ & $0.92$\\
    W0C1L100    & $\mathbf{\tilde{C}}^\text{F+V\hphantom{D}}_{9}$  & $500$ & $1$  & sqrt & $0.96$ & $0.97$ & $0.98$ & $0.97$ & $0.96$ & $0.97$ & $0.96$ & $0.96$\\
    W0C5L25     & $\mathbf{\tilde{C}}^\text{C+V\hphantom{D}}_{5}$  & $100$ & $10$ & sqrt & $0.85$ & $0.74$ & $0.84$ & $0.81$ & $0.81$ & $0.68$ & $0.81$ & $0.77$\\
    W0C5L50     & $\mathbf{\tilde{C}}^\text{SD+V}_{9}$             & $500$ & $10$ & sqrt & $0.84$ & $0.76$ & $0.86$ & $0.82$ & $0.81$ & $0.71$ & $0.82$ & $0.78$\\
    W0C5L100    & $\mathbf{\tilde{C}}^\text{F+V\hphantom{D}}_{9}$  & $500$ & $1$  & sqrt & $0.90$ & $0.90$ & $0.94$ & $0.91$ & $0.86$ & $0.87$ & $0.92$ & $0.88$\\
    W0C10L25    & $\mathbf{\tilde{C}}^\text{C+V\hphantom{D}}_{7}$  & $500$ & $10$ & sqrt & $0.79$ & $0.67$ & $0.74$ & $0.73$ & $0.77$ & $0.61$ & $0.69$ & $0.69$\\
    W0C10L50    & $\mathbf{\tilde{C}}^\text{C+V\hphantom{D}}_{9}$  & $100$ & $10$ & sqrt & $0.76$ & $0.70$ & $0.72$ & $0.73$ & $0.74$ & $0.64$ & $0.70$ & $0.69$\\
    W0C10L100   & $\mathbf{\tilde{C}}^\text{LC\hphantom{+V}}_{99}$            & $100$ & $10$ & sqrt & $0.88$ & $0.76$ & $0.91$ & $0.85$ & $0.84$ & $0.73$ & $0.89$ & $0.82$\\
    W1C0        & $\mathbf{\tilde{C}}^\text{SD+V}_{9}$             & $500$ & $1$  & sqrt & $0.96$ & $0.95$ & $0.98$ & $0.97$ & $0.94$ & $0.95$ & $0.97$ & $0.95$\\
    W1C1L25     & $\mathbf{\tilde{C}}^\text{SD+V}_{9}$             & $500$ & $1$  & sqrt & $0.90$ & $0.89$ & $0.96$ & $0.92$ & $0.86$ & $0.87$ & $0.94$ & $0.89$\\
    W1C1L50     & $\mathbf{\tilde{C}}^\text{SD+V}_{9}$             & $500$ & $1$  & sqrt & $0.92$ & $0.91$ & $0.97$ & $0.93$ & $0.91$ & $0.91$ & $0.95$ & $0.92$\\
    W1C1L100    & $\mathbf{\tilde{C}}^\text{F+V\hphantom{D}}_{9}$  & $500$ & $1$  & sqrt & $0.95$ & $0.95$ & $0.98$ & $0.96$ & $0.95$ & $0.93$ & $0.96$ & $0.95$\\
    W1C5L25     & $\mathbf{\tilde{C}}^\text{F+V\hphantom{D}}_{7}$  & $500$ & $10$ & sqrt & $0.85$ & $0.73$ & $0.85$ & $0.81$ & $0.83$ & $0.67$ & $0.81$ & $0.77$\\
    W1C5L50     & $\mathbf{\tilde{C}}^\text{SD+V}_{9}$             & $500$ & $10$ & None & $0.83$ & $0.76$ & $0.85$ & $0.81$ & $0.81$ & $0.72$ & $0.81$ & $0.78$\\
    W1C5L100    & $\mathbf{\tilde{C}}^\text{F+V\hphantom{D}}_{9}$  & $500$ & $1$  & sqrt & $0.90$ & $0.89$ & $0.92$ & $0.90$ & $0.85$ & $0.86$ & $0.89$ & $0.87$\\
    W1C10L25    & $\mathbf{\tilde{C}}^\text{F+V\hphantom{D}}_{9}$  & $500$ & $10$ & sqrt & $0.79$ & $0.67$ & $0.73$ & $0.73$ & $0.78$ & $0.61$ & $0.69$ & $0.69$\\
    W1C10L50    & $\mathbf{\tilde{C}}^\text{SD+V}_{9}$             & $500$ & $10$ & sqrt & $0.77$ & $0.69$ & $0.74$ & $0.73$ & $0.74$ & $0.64$ & $0.68$ & $0.69$\\
    W1C10L100   & $\mathbf{\tilde{C}}^\text{SD+V}_{9}$             & $100$ & $10$ & None & $0.87$ & $0.80$ & $0.87$ & $0.84$ & $0.83$ & $0.75$ & $0.82$ & $0.80$\\
    W10C0       & $\mathbf{\tilde{C}}^\text{SD+V}_{8}$             & $500$ & $1$  & sqrt & $0.91$ & $0.89$ & $0.96$ & $0.92$ & $0.89$ & $0.87$ & $0.92$ & $0.90$\\
    W10C1L25    & $\mathbf{\tilde{C}}^\text{SD+V}_{4}$             & $500$ & $10$ & sqrt & $0.92$ & $0.82$ & $0.96$ & $0.90$ & $0.89$ & $0.79$ & $0.92$ & $0.87$\\
    W10C1L50    & $\mathbf{\tilde{C}}^\text{SD+V}_{8}$             & $100$ & $1$  & sqrt & $0.89$ & $0.87$ & $0.94$ & $0.90$ & $0.86$ & $0.85$ & $0.92$ & $0.88$\\
    W10C1L100   & $\mathbf{\tilde{C}}^\text{SD+V}_{8}$             & $500$ & $1$  & sqrt & $0.89$ & $0.88$ & $0.95$ & $0.91$ & $0.88$ & $0.86$ & $0.92$ & $0.89$\\
    W10C5L25    & $\mathbf{\tilde{C}}^\text{C+V\hphantom{D}}_{5}$  & $500$ & $10$ & sqrt & $0.86$ & $0.73$ & $0.85$ & $0.81$ & $0.83$ & $0.67$ & $0.79$ & $0.76$\\
    W10C5L50    & $\mathbf{\tilde{C}}^\text{SD+V}_{9}$             & $500$ & $10$ & sqrt & $0.84$ & $0.76$ & $0.84$ & $0.81$ & $0.82$ & $0.70$ & $0.81$ & $0.78$\\
    W10C5L100   & $\mathbf{\tilde{C}}^\text{SD+V}_{8}$             & $100$ & $10$ & sqrt & $0.87$ & $0.80$ & $0.91$ & $0.86$ & $0.84$ & $0.76$ & $0.87$ & $0.83$\\
    W10C10L25   & $\mathbf{\tilde{C}}^\text{F+V\hphantom{D}}_{8}$  & $100$ & $10$ & sqrt & $0.77$ & $0.68$ & $0.73$ & $0.72$ & $0.75$ & $0.62$ & $0.69$ & $0.69$\\
    W10C10L50   & $\mathbf{\tilde{C}}^\text{SD+V}_{7}$             & $100$ & $10$ & sqrt & $0.76$ & $0.70$ & $0.74$ & $0.74$ & $0.75$ & $0.64$ & $0.69$ & $0.69$\\
    W10C10L100  & $\mathbf{\tilde{C}}^\text{SD+V}_{8}$             & $500$ & $10$ & None & $0.84$ & $0.75$ & $0.85$ & $0.81$ & $0.82$ & $0.71$ & $0.81$ & $0.78$\\
    W100C0      & $\mathbf{\tilde{C}}^\text{F+V\hphantom{D}}_{8}$  & $500$ & $10$ & sqrt & $0.89$ & $0.74$ & $0.81$ & $0.81$ & $0.88$ & $0.69$ & $0.78$ & $0.78$\\
    W100C1L25   & $\mathbf{\tilde{C}}^\text{F+V\hphantom{D}}_{8}$  & $500$ & $10$ & None & $0.87$ & $0.74$ & $0.82$ & $0.81$ & $0.85$ & $0.70$ & $0.77$ & $0.77$\\
    W100C1L50   & $\mathbf{\tilde{C}}^\text{SD+V}_{4}$             & $500$ & $10$ & sqrt & $0.90$ & $0.74$ & $0.81$ & $0.82$ & $0.86$ & $0.68$ & $0.76$ & $0.77$\\
    W100C1L100  & $\mathbf{\tilde{C}}^\text{F+V\hphantom{D}}_{8}$  & $100$ & $10$ & sqrt & $0.89$ & $0.74$ & $0.81$ & $0.81$ & $0.87$ & $0.69$ & $0.77$ & $0.78$\\
    W100C5L25   & $\mathbf{\tilde{C}}^\text{C+V\hphantom{D}}_{5}$  & $100$ & $10$ & None & $0.84$ & $0.71$ & $0.78$ & $0.77$ & $0.81$ & $0.65$ & $0.73$ & $0.73$\\
    W100C5L50   & $\mathbf{\tilde{C}}^\text{SD+V}_{7}$             & $100$ & $10$ & sqrt & $0.82$ & $0.72$ & $0.76$ & $0.77$ & $0.79$ & $0.67$ & $0.73$ & $0.73$\\
    W100C5L100  & $\mathbf{\tilde{C}}^\text{C+V\hphantom{D}}_{7}$  & $500$ & $10$ & None & $0.82$ & $0.73$ & $0.79$ & $0.78$ & $0.79$ & $0.68$ & $0.75$ & $0.74$\\
    W100C10L25  & $\mathbf{\tilde{C}}^\text{F+V\hphantom{D}}_{7}$  & $500$ & $10$ & sqrt & $0.79$ & $0.68$ & $0.68$ & $0.72$ & $0.77$ & $0.60$ & $0.64$ & $0.67$\\
    W100C10L50  & $\mathbf{\tilde{C}}^\text{C+V\hphantom{D}}_{7}$  & $100$ & $10$ & None & $0.78$ & $0.69$ & $0.68$ & $0.72$ & $0.76$ & $0.62$ & $0.65$ & $0.68$\\
    W100C10L100 & $\mathbf{\tilde{C}}^\text{F+V\hphantom{D}}_{8}$  & $500$ & $10$ & sqrt & $0.80$ & $0.71$ & $0.75$ & $0.75$ & $0.78$ & $0.66$ & $0.70$ & $0.72$\\
    \hline
    \end{tabular}
    }
    \tablefoot{We also provide the validation class recalls for the dark companion ($R^\text{V}_\text{DC}$), semidetached ($R^\text{V}_\text{SD}$), and contact ($R^\text{V}_\text{C}$) binary light curves as well as the test class recalls ($R^\text{T}_\text{DC}$, $R^\text{T}_\text{SD}$, $R^\text{T}_\text{C}$) and the test macro recall ($R^\text{T}_\text{M}$) of the best performing classifiers. See Sect.~\ref{sec:data} for the description of the synthetic samples and Sect.~\ref{sec:random_forests_macro_recall_methods} for the definitions of the augmented representations and the hyperparameters. The symbol $\mathbf{\tilde{C}}^\text{LC}_{99}$ denotes the extended full representation.}
    \end{center}
\end{table*}

\begin{figure*}[ht!]
    \centering
    \begin{subfigure}[b]{0.95\textwidth}
        \centering
        \includegraphics[width=\textwidth]{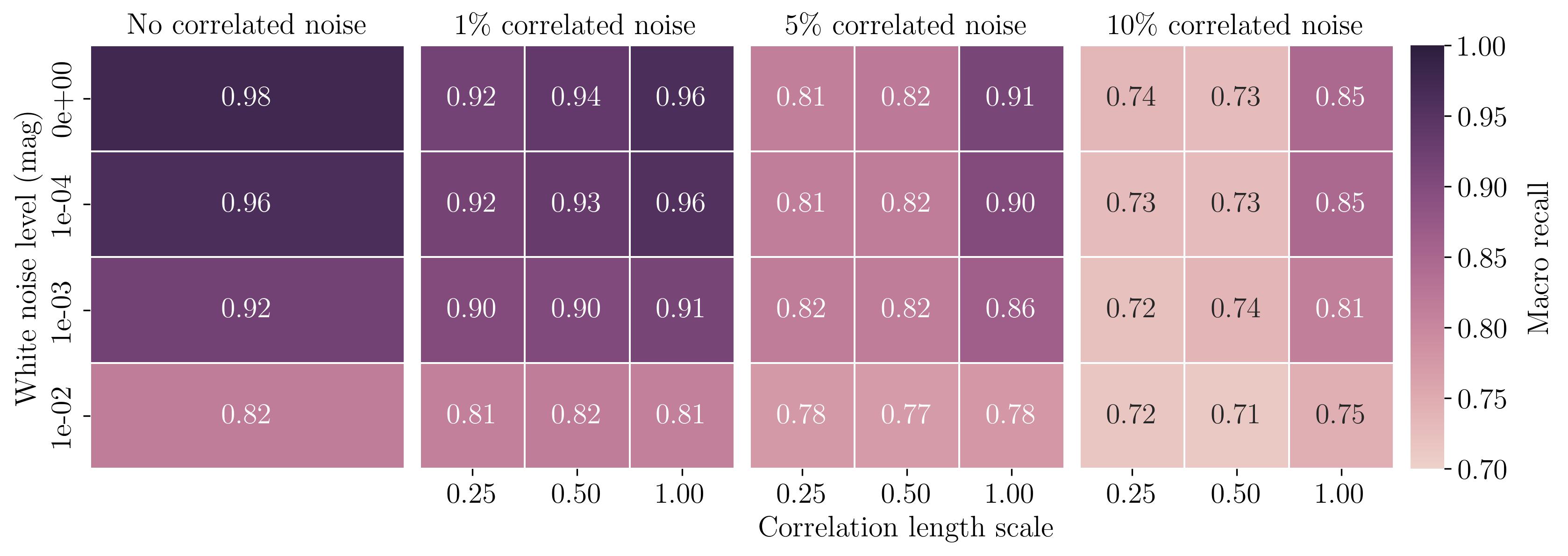}
    \end{subfigure}
    \\
    \begin{subfigure}[b]{0.95\textwidth}
        \centering
        \includegraphics[width=\textwidth]{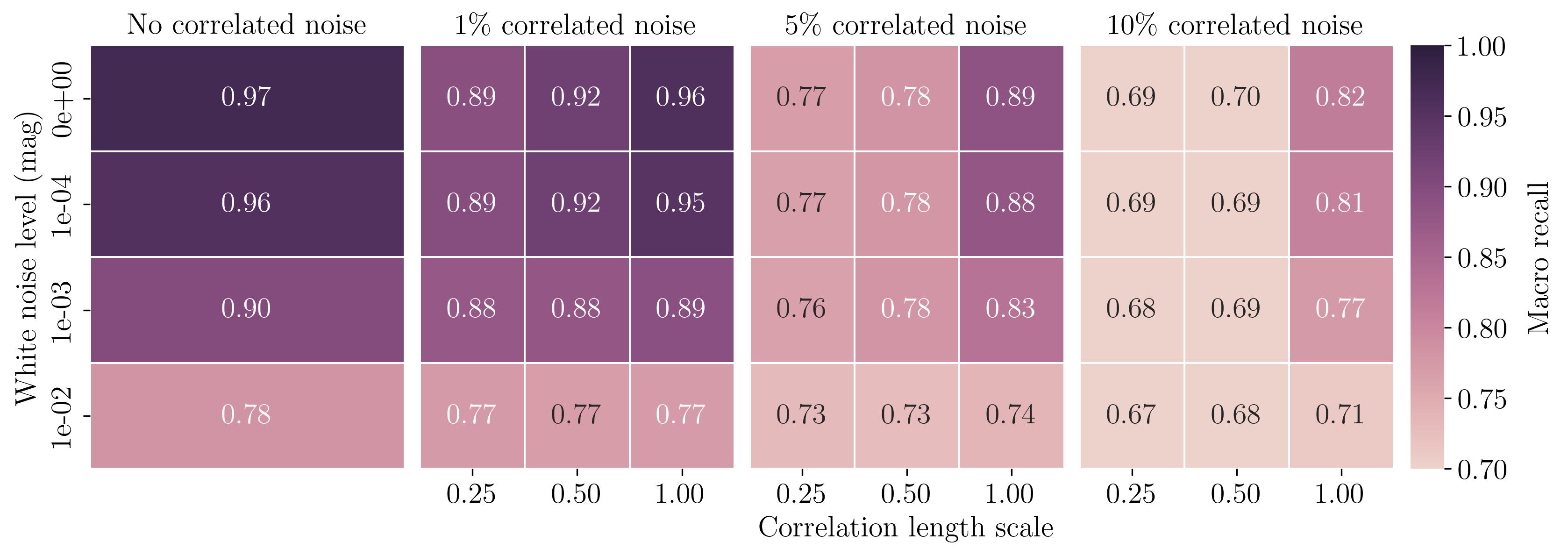}
    \end{subfigure}
    \caption{
    Best validation macro recalls achieved by the random forest classifiers trained on the latent representations of the synthetic samples (top panel) and the macro recalls of the best performing classifiers evaluated on the test sets of the synthetic samples (bottom panel).
    \label{fig:validation_test_macro_recalls}}
\end{figure*}

\begin{figure*}[ht!]
    \centering
    \includegraphics[width=0.95\textwidth]{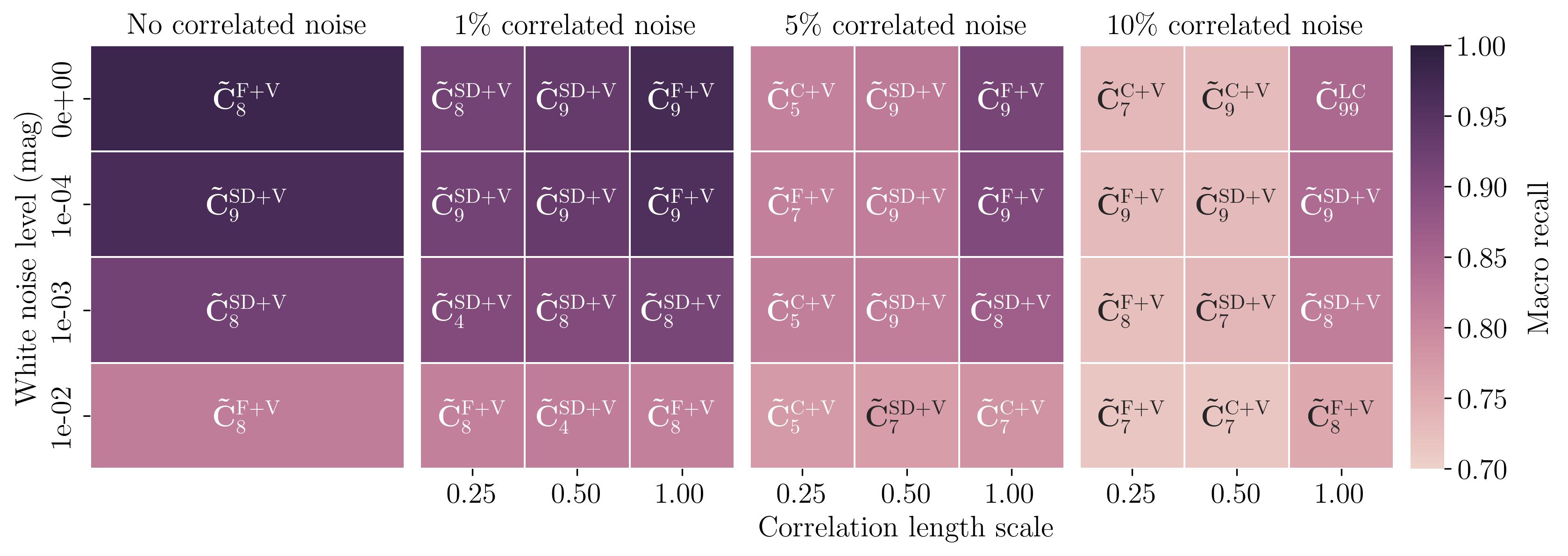}
    \caption{
    Augmented latent representations that yielded the best macro recalls on the validation sets of the synthetic samples. See Sects.~\ref{sec:data} and \ref{sec:random_forests_macro_recall_methods} for the descriptions of the synthetic samples and the augmented representations, respectively. The symbol $\mathbf{\tilde{C}}^\text{LC}_{99}$ denotes the extended full representation.
    \label{fig:augmented_representations_and_dimensions_at_best_recalls}}
\end{figure*}

\begin{figure*}[ht!]
    \centering
    \begin{subfigure}[b]{0.95\textwidth}
        \centering
        \includegraphics[width=\textwidth]{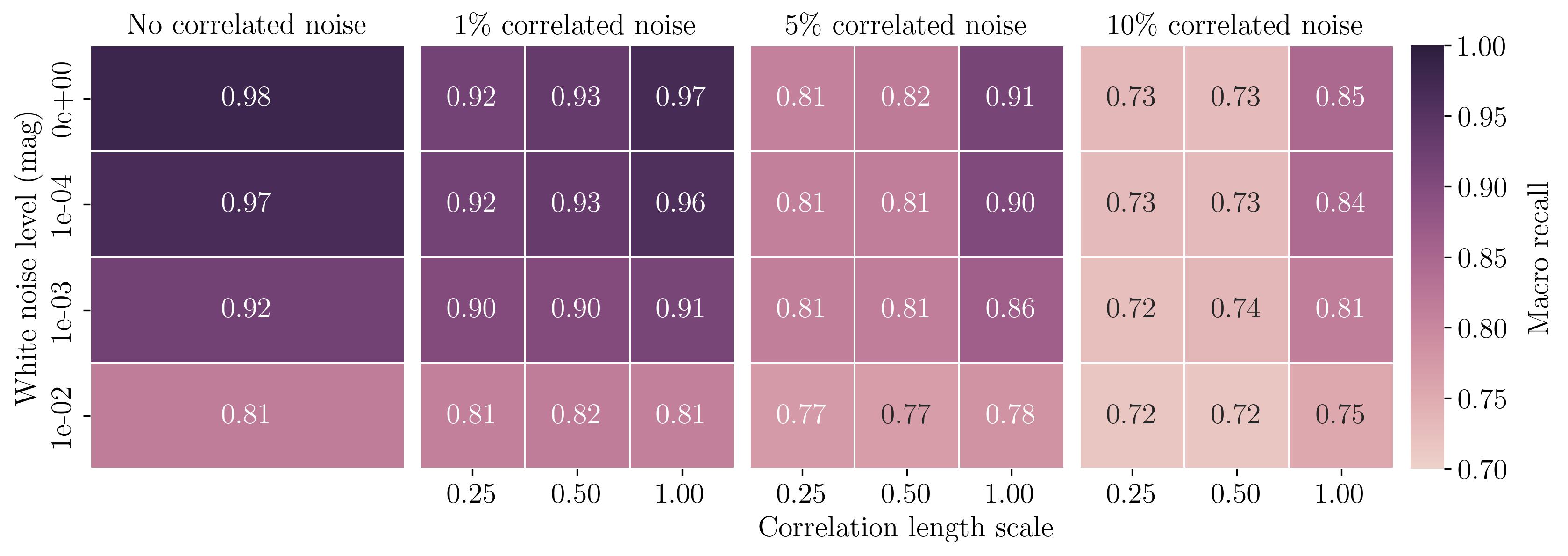}
    \end{subfigure}
    \\
    \begin{subfigure}[b]{0.95\textwidth}
        \centering
        \includegraphics[width=\textwidth]{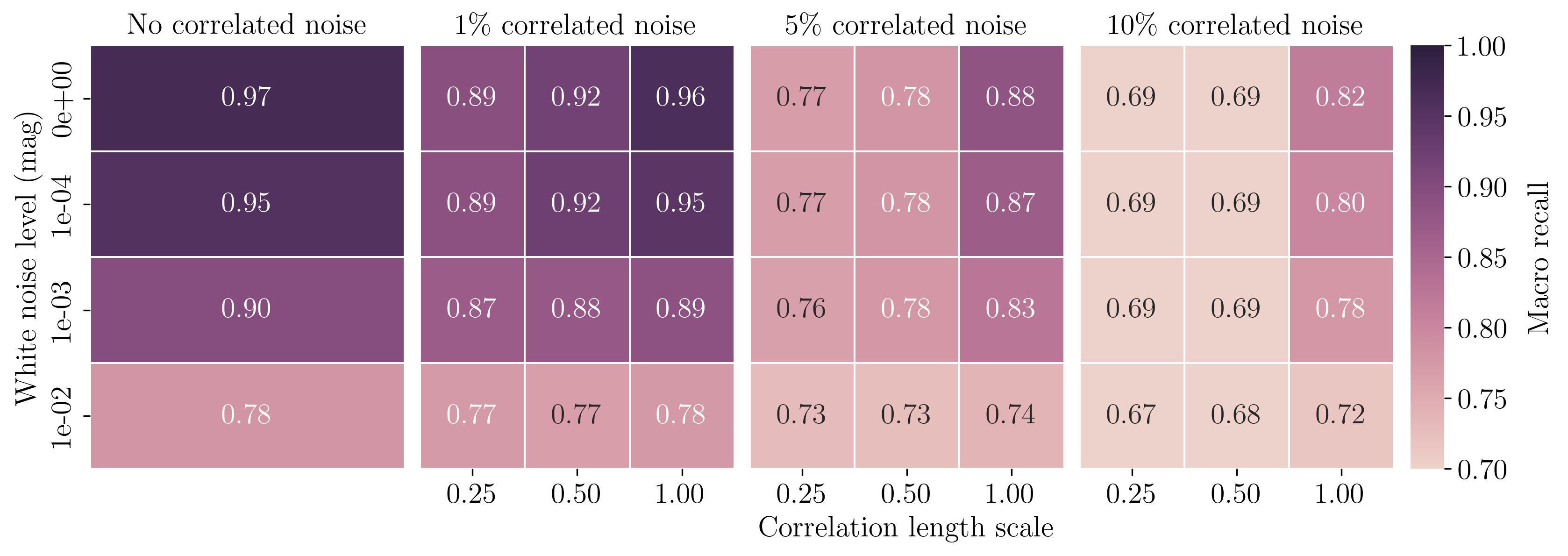}
    \end{subfigure}
    \caption{
    Highest validation macro recalls achieved by the random forest classifiers trained on the augmented latent representations of the synthetic samples (top panel) and the macro recalls of the best performing classifiers evaluated on the test sets of the synthetic samples (bottom panel).
    \label{fig:validation_test_macro_recalls_augmented_representations}}
\end{figure*}

\end{appendix}
\end{document}